\documentclass[preprint]{aastex}

\usepackage{hyperref}
\usepackage{graphicx}
\usepackage{color}
\graphicspath{{/Plots/}}
\usepackage{natbib}
\begin{document}

\title{The Structure of the Milky Way's Hot Gas Halo}
\author{Matthew J. Miller, Joel N. Bregman}
\affil{Department of Astronomy, University of Michigan, Ann Arbor, MI 48104}
\email{mjmil@umich.edu, jbregman@umich.edu}

\begin{abstract}

The Milky Way's million degree gaseous halo contains a considerable amount of mass that, depending on its structural properties, can be a significant mass component. In order to analyze the structure of the Galactic halo, we use \textit{XMM-Newton} Reflection Grating Spectrometer archival data and measure O {\scriptsize{VII}} K$\alpha$ absorption-line strengths toward 26 active galactic nuclei, LMC X-3, and two Galactic sources (4U 1820-30 and X1735-444).  We assume a $\beta$-model as the underlying gas density profile and find best-fit parameters of $n_{\circ}$ = 0.46 $^{+0.74}_{-0.35}$ cm$^{-3}$, $r_c$ = 0.35 $^{+0.29}_{-0.27}$ kpc, and $\beta$ = 0.71 $^{+0.13}_{-0.14}$.  These parameters result in halo masses ranging between $M$(18 kpc) = 7.5 $^{+22.0}_{-4.6}$ $\times$ 10$^8$ $M_\sun$ and $M$(200 kpc) = 3.8 $^{+6.0}_{-0.5}$ $\times$ 10$^{10}$ $M_\sun$ assuming a gas metallicity of $Z$ = 0.3 $Z_{\odot}$, which are consistent with current theoretical and observational work.  The maximum baryon fraction from our halo model of $f_b$ = 0.07 $^{+0.03}_{-0.01}$ is significantly smaller than the universal value of $f_b$ = 0.171, implying the mass contained in the Galactic halo accounts for 10\% - 50\% of the missing baryons in the Milky Way.  We also discuss our model in the context of several Milky Way observables, including ram pressure stripping in dwarf spheroidal galaxies, the observed X-ray emission measure in the 0.5 - 2 keV band, the Milky Way's star formation rate, spatial and thermal properties of cooler gas ($\sim$$10^5$ K) and the observed \textit{Fermi} bubbles toward the Galactic center.  Although the metallicity of the halo gas is a large uncertainty in our analysis, we place a lower limit on the halo gas between the Sun and the Large Magellanic Cloud (LMC).  We find that $Z$ $\gtrsim$ 0.2 $Z_{\odot}$ based on the pulsar dispersion measure toward the LMC.  

\end{abstract}

\section{Introduction}

\ The detection of hot gas at zero redshift by \textit{Chandra, XMM-Newton}, and \textit{FUSE} implies there exists a reservoir of gas that potentially contains a significant amount of baryonic mass in the Milky Way.  However, there have only been detections of this hot Galactic halo with little detailed analysis of its structural properties using multiple sightlines until this point.  If the density profile of this halo was constrained, the mass could be calculated and compared to the other baryon mass components of the Milky Way in an attempt to account for some or all of the ``missing baryons" in the local universe.  

\ There has been extensive work in detecting and analyzing hot gaseous halos in galaxies other than the Milky Way as probes of galaxy formation and evolution.  Detailed analyses of individual galaxies have revealed that other galaxies have considerable amounts of mass in their extended gaseous halos, but there is not enough to account for their missing baryons \citep{breghouck97,li08,ab10,ab11}.  In addition to detailed analyses of individual galaxies, there have been numerous studies of the global properties of diffuse X-ray emission around galaxies \citep{opc03,strickland04,tullmann06}.  These properties offer a foundation for comparing the Milky Way's gaseous halo to other galaxies.  

\ The primary tracers of the Milky Way's hot gas halo are O {\scriptsize{VII}} and O {\scriptsize{VIII}} that exist in the 10$^6$ - 10$^7$ K range \citep{pk03}.  These X-ray lines have primarily been observed in absorption of active galactic nuclei (AGNs) and blazar spectra \citep{nicastro02,rasmussen03,wang05,williams05,fang06,yaowang07,hag10} as well as Galactic X-ray binary spectra \citep{yaowang05,hag11} using \textit{Chandra} and \textit{XMM-Newton}.  The lines have also been observed in emission with the Diffuse X-Ray Spectrometer by \cite{mccammon02}.  In addition to this X-ray emitting/absorbing gas, O {\scriptsize{VI}}, the most common ion in $\sim$10$^{5.5}$ K gas, has been extensively studied with \textit{FUSE} \citep{sembach03,wakker03,fox04}.  While the interaction between these gas phases is an area of interest, mass estimates from this cooler gas have not helped solve the missing baryon problem.  

\ The structural properties of the Galactic halo have not been well constrained up until this point.  The combination of O {\scriptsize{VII}} emission and absorption has constrained the temperature of the halo to be between log T = 6.1 - 6.4, however these estimates come primarily from analyses of single lines of sight toward the highest signal-to-noise (S/N) targets \citep{wang05,yaowang07,hag10}.  Attempts to constrain the density profile and thus the mass of the halo have been limited by the number of extragalactic lines of sight with high enough S/N \citep{bld07,gupta12} or by only using Galactic sources to constrain the density profile \citep{yaowang05}.  If the halo extends past the disk of the galaxy, the density profile could only be constrained with multiple extragalactic sightlines.

\ \cite{bld07} studied O {\scriptsize{VII}} absorption lines using the Reflection Grating Spectrometer (RGS) on \textit{XMM-Newton} with the goal to determine if these zero redshift absorption lines were from gas associated with the Milky Way's halo or from a Local Group medium.  This size discrepancy between these two scenarios is critical in terms of estimating the baryon mass of gas at this temperature.  Their analysis indicated the equivalent widths of O {\scriptsize{VII}} lines are positively correlated with angle toward the Galactic center as opposed to M31 (toward the center of mass of the Local Group).  This implies these lines originate in a gas reservoir confined to the Milky Way as opposed to the Local Group.  

\ In this paper, we expand upon \cite{bld07} by adding three additional targets to our sample (one AGN and two Galactic targets) as well as provide a more detailed analysis of the structure and global properties of the hot Galactic halo, specifically the density profile.  We use O {\scriptsize{VII}} column densities along different sightlines throughout the Milky Way to determine the density profile of the hot Galactic halo.  We then estimate the mass contained in the halo as well as compare our density profile with numerous observational constraints.  The outline of the paper is as follows.  In Section 2, we present our object selection and data analysis.  In Section 3, we discuss our model fitting procedure as well as the different models we consider.  In Section 4, we examine the consistency of our model with previously established constraints as well as compare our hot Galactic halo to different phases of the interstellar medium (ISM).

\section{Object Selection}

\ Our initial target list was identical to that of \cite{bld07}, which was constructed from the brightest AGNs in the \textit{XMM-Newton} RGS archival data.  In addition to finding sources with suitable X-ray fluxes at 21.6 $\AA$, the goal was also to find sources with O {\scriptsize{VII}} equivalent width uncertainties less than 10 m{$\AA$} (for typical equivalent widths of about 20 m$\AA$).  This resulted in a sample of 25 AGNs plus an additional source in the Large Magellanic Cloud (LMC).  

\ In addition to the original sample, we added an additional AGN (ESO 141-G055) and two Galactic low-mass X-ray binaries (LMXBs; 4U 1820-30, X1735-444) to our target list (Table 1).  These added source spectra can be seen in Figure 1 while the distribution of our sources on the sky can be seen in Figure 2.  The motivation for adding these sources is their proximity to the Galactic center.  Two of the parameters of interest in our model fitting are sensitive to small galactocentric radii ($r \lesssim$ 2 kpc), which would not be probed by the original sample (see Figure 2).  The inclusion of these three sources allowed us to analyze the central region of the density profile better than the previous target list would have allowed.  The additional targets also have small uncertainties in the measured equivalent widths ($\lesssim$ 25\%) compared to the rest of the sample.  

\ We are unable to distinguish between O {\scriptsize{VII}} halo gas absorption and intrinsic absorption for our Galactic targets due to the resolution of the RGS at 21.6 $\AA$.  However, the X-ray spectra of our two Galactic sources have been previously observed with no indication of intrinsic O {\scriptsize{VII}} absorption.  There has been evidence for a photoionized wind from an accretion disk in 4U 1820-30 \citep{costantini12}.  However this wind has only been detected in the lower ionization states of oxygen (O {\scriptsize{V}} and O {\scriptsize{VI}}) and not in the O {\scriptsize{VII}} absorption lines used in this study.  \cite{yaowang05} also analyzed X-ray spectra for a sample of Galactic sources, including X1735-444, to constrain the structure of the local hot ISM.  They examine the possibility that some of the unresolved O {\scriptsize{VII}} absorption may come from photoionized winds, thus contaminating absorption from halo gas.  However, they conclude this scenario is unlikely since the calculated radii where the lines would be produced are larger than the measured binary separations in six out of seven targets.  We assume throughout our model fitting procedure that the O {\scriptsize{VII}} lines are entirely due to the halo gas and not associated with our LMXB targets.

\ The data reduction and data analysis for our sample was identical to that of \cite{bld07} since this work is an extension of their results.  Thus, we refer the reader to the aforementioned paper for any details concerning the data acquisition, methods of fitting the spectra, or determining equivalent widths.

\section{Model Fitting}

\ Our model fitting procedure focuses on the comparison between measured electron column densities of our targets and theoretical column densities calculated by our model density profile.  We consider both spherical and flattened density models in our model fitting procedure.  The spherical profile is the simplest model we consider in the fitting process while the flattened profile adds one additional free parameter to account for the potential disk-like shape of the gas distribution.  The coordinate transformations from a galactocentric density profile to a Sun-centered line of sight distance profile are

	\begin{equation}
		R^2 = R_o^2 + d^2cos(b)^2 - 2dR_ocos(b)cos(l)
	\end{equation}
	\begin{equation}
		z^2 = d^2sin(b)^2
	\end{equation}
	\begin{equation}
		r^2 = R^2 + z^2
	\end{equation}

\noindent In these coordinates, $d$ is the line of sight distance, $b$ and $l$ are Galactic latitude and longitude, respectively, $R_o$ is the distance from the Sun to the Galactic center (we assume 8.5 kpc throughout our analysis), and $r$ is the galactocentric radius.  We constrain the density profile of the halo by generating model column densities for a given set of model parameters and finding the parameter set that minimizes the $\chi^2$ calculated from our data.

\subsection{Column Density Calculation}

\ In order to convert from measured O {\scriptsize{VII}} equivalent widths to total electron column densities, we follow several assumptions presented in \cite{bld07}.  We initially convert the equivalent widths to column densities assuming the lines are optically thin.  In this case, the resulting linear conversion between the measured equivalent width and O {\scriptsize{VII}} column density is $N$(O {\scriptsize{VII}}) = 3.48 $\times$ 10$^{14}$ EW, where EW is the O {\scriptsize{VII}} equivalent width in m{$\AA$} and the column density has units of cm$^{-2}$.  However, recent work has shown the lines are likely mildly saturated based on the observed K$\beta$ to K$\alpha$ ratio for the O {\scriptsize{VII}} ion \citep{williams05, gupta12}.  To analyze the saturation effects in our lines, we make assumptions on the Doppler widths of the lines since the lines are not resolved by the RGS.  We expect the gas to be turbulent from supernovae mixing and subject to turbulent mixing layers between the hotter and cooler phases of halo gas \citep{begelman90, kwak10}.  Thus, we assume the gas is turbulent at the sound speed of hydrogen, which is consistent with simulations of halo gas \citep{fp06, cen11}.  This results in an assumed Doppler width of $b$ $\approx$ 150 km s$^{-1}$.  For a typical equivalent width in our sample ($\approx$ 20 m$\AA$) the optical depth at line center is of order unity, implying minor saturation corrections for our lines. To account for this, we initially calculate column densities linearly from our measured equivalent widths and determine correction factors using the curve of growth and assuming a Doppler width associated with turbulent motions discussed above.  We use both sets of column densities to calculate best-fit parameters for our halo model due to the uncertainty in the Doppler widths of the lines and to examine how our best-fit parameters change after accounting for line saturation (see Section 4.4).  However, we use the model fitting results from the saturated column densities in our analysis since the lines are expected to be minorly saturated.  Both sets of column densities can be found in Table 1.

\ We make several assumptions while converting between O {\scriptsize{VII}} and total electron column densities.  We assume that the abundance of oxygen is similar to the solar value and thus adopt a value of log($A_O$) = 8.74  \citep{hol01}. This results in a total electron column of

	\begin{equation}
	 N_e = 4.4 \times 10^{19} \left(\frac{N_{\textrm{O \scriptsize{VII}}}}{10^{16} \textrm{ cm}^{-2}} \right) \left( \frac{f}{0.5} \right)^{-1} \left( \frac{Z}{Z_{\odot}} \right)^{-1}\textrm{ cm}^{-2}
	\end{equation}

\noindent where $f$ is the ion fraction of O {\scriptsize{VII}} and $Z$ is the metallicity of the gas.  We assume an ion fraction of 0.5 for O {\scriptsize{VII}}, which is constrained by the temperature of the gas.  We note that while we do assume a solar gas metallicity initially, the metallicity of the halo gas is a significant uncertainty in our analysis.  We expect a portion of the halo to be enriched due to feedback from the Milky Way disk, but the true metallicity is likely less than the solar value.  We discuss the implications of this assumption throughout our analysis.

\subsection{Spherical Model}

\ The simplest model we consider in our analysis is the  $\beta$-model.  We choose this model as opposed to a simpler model (such as a uniform density sphere) because it reproduces the observed X-ray surface brightness profiles of other galaxies \citep{opc03}.  The $\beta$-model consists of three parameters and is defined as

	\begin{equation}
	 n(r) = {n_o}({1 + ({r}/{r_c})^2})^{-{3\beta}/{2}}
	\end{equation}

\noindent In this model, $n_\circ$ is the core density, $r_c$ is the core radius, and $\beta$ is the slope of the profile at large radii.  The parameters $n_\circ$ and $r_c$ describe the density near the center of the profile since $n_\circ$ is the density at $r$ = 0 and typical values for $r_c$ are $\lesssim$ 1 kpc.  These parameters are of little importance for mass estimates of the halo gas since majority of the mass comes from material at large radii.  On the other hand, $\beta$ defines the behavior of the density profile at $r$ $\textgreater$  $r_c$ and is thus the primary parameter of interest.  

\ Our best-fit parameters can be seen in Table 2 while the O {\scriptsize{VII}} column densities resulting from our best-fit models can be seen in Table 1.  We also show how our observed saturated column densities compare with our best-fit model column densities in Figure 3.  Initially, the best-fit model results in a $\chi^2$ that is unacceptably large in both the optically thin and saturated column density conversions ($\chi_{thin}^2$(dof) = 85.2 (26), $\chi_{saturated}^2$(dof) = 56.3 (26)), which is possible if the variation due to substructure of the absorbing medium is larger than the statistical uncertainty of the equivalent widths.  In order to account for this intrinsic variation, we add an additional uncertainty to all the equivalent widths of 7.5 m$\AA$\ for the optically thin column density conversion and 7.2 m$\AA$ in the saturated case ($\sim$30\% of the average equivalent width).  We also examine the variation in our calculated $\chi^2$ to place a constraint on the halo size.  Initially, we assume a halo size of 200 kpc when determining our best-fit parameters. By changing the size of the halo after we find our best-fit parameters, we can determine how the halo size changes the minimum $\chi^2$ until our fit becomes unacceptable.  We find a halo size 32 kpc at the 95\%  confidence level and 18 kpc at the 99\%  confidence level for both column density calculations.

\ The quality of the constraints on the halo parameters depends on both the quality of the data used as well as the location of our sources on the sky.  The parameters $n_\circ$ and $r_c$ are the least constrained primarily due to only three of our targets passing near the Galactic center.  In particular, the two Galactic targets that come closest to the Galactic center are less than 8.5 kpc away and thus do not probe the full inner region of the halo.  Note that this would not be the case if $r_c$ were larger, implying a more extended profile.  The fact that these two parameters are not well constrained results from a degeneracy between them which is most apparent in Figure 4.  To account for this, we note that for $r_c$ $\ll$ $r$ Equation (1) can be approximated by the power law

	\begin{equation}
	 n(r) \approx \frac{n_or_c^{3\beta}}{r^{3\beta}}
	\end{equation}

\noindent This reduces the dimensionality of the problem by making the free parameters a constant ($n_or_c^{3\beta}$) and $\beta$.  We expect the constraints on these parameters to be more reliable than the three parameter model due to the few lines of sight near the Galactic center.  The 1$\sigma$, 2$\sigma$, and 3$\sigma$ contours for all of our parameter spaces can be seen in Figure 4, considering the three and two parameter spherical models and using saturated column densities.  We note little difference in the quality of our constraints between the three and two parameter spherical models.  Fortunately, the parameter $\beta$ is relatively well constrained for both models due to the majority of our sample being extragalactic targets.

\subsection{Flattened Model}

\ We also consider a flattened model in our fitting process by modifying the spherical $\beta$-model.  The motivation behind considering a flattened density profile is based on the angular momentum associated with the Milky Way.  The rotation of dark matter and baryons in the Milky Way could potentially alter a spherical gas profile into a flattened profile.  We modify the traditional $\beta$-model in the following way:
	\begin{equation}
	 n(r) = {n_o}({1 + ({R}/{R_c})^2 + ({z}/{z_c})^2})^{-{3\beta}/{2}}
	\end{equation}

\noindent In this flattened model, $R$ is the radius in the plane of the disk and $z$ is the height off of the Galactic plane with $R_c$ and $z_c$ representing the effective core radii for each coordinate.  The parameters $R_c$ and $z_c$ are thus measures of density profile flattening.  For this portion of our model fitting procedure, we fix the central density and slope of the density profile to their best-fit values associated with the spherical model and saturated column densities.  Thus, we consider a flattened model only to see how the core radii associated with the orthogonal coordinates $R$ and $z$ change from a single, spherical core radius.  

\ The best-fit parameters for our flattened model can be seen in Table 2.  The flattened model initially has the same issue as the spherical model in that the best-fit parameters result in a $\chi^2$ that is unacceptably large.  Thus, we adopt the same procedure applied to the spherical model and add an additional uncertainty to the equivalent widths along each line of sight.  We find that we must add the same additional uncertainty as the spherical case, 7.2 m{$\AA$}, to the equivalent widths to obtain an acceptable $\chi^2$.  After accounting for this additional uncertainty, the best-fit parameters still result in core radii of less than 1 kpc and an axial ratio of $\sim$3/2.  The 1$\sigma$, 2$\sigma$, and 3$\sigma$ contours for the core radii parameters can be seen in Figure 5 and indicate that profile is preferentially elongated in the $R$ direction.  However, the $\chi^2$ change is small between the spherical and flattened models.  This implies that the flattened and spherical models are equivalent in describing the density profile of the halo.  Thus, we assume the density profile is spherical for the rest of our analysis for simplicity.

\subsection{Negative Column Densities}

\ Our sample contains three negative equivalent width measurements with corresponding negative column density conversions (see Table 1), which are possible if the S/N (the ratio between the measured equivalent width and corresponding uncertainty) is not sufficiently high for the observations.  Although negative columns are not physical, we emphasize our measurements are all consistent with positive values based on their 1$\sigma$ uncertainties and our model fitting procedure is sensitive to the difference between the measured and model column densities and the uncertainties associated with our measurements.  Furthermore, the negative equivalent width measurements are not heavily weighted in our model fitting procedure due to their large uncertainties.  We explore the effects of negative equivalent width measurements on our model fitting procedure by truncating the negative measurements at zero and refitting the data assuming the same spherical model and saturation effects discussed above.  The best-fit parameters for our sample with these truncated values are nearly identical to our results from our original sample.  Alternatively, we apply a S/N cut to our sample to remove the negative equivalent width measurements from our model fitting procedure and determine how our results change.  We choose a S/N threshold of 1.1, which removes the three negative equivalent width measurements as well as three positive measurements.  The model fitting results from these two altered samples can be seen in Table 2.  The best-fit parameters from these samples are consistent with our initial model fitting results based on their 1$\sigma$ uncertainties.  In particular, the parameter $\beta$ is nearly identical between our initial and truncated sample results and shows a 16\% variation between the initial and S/N cut sample results.  It should be noted that both of these alterations to our sample impose a slight bias to our results toward the higher S/N observations, which may be due to strong absorption features or simply low noise measurements.  However, our model fitting results are not strongly dependent to these changes due to the weak weighting of the low S/N measurements in our initial model fitting procedure.  We therefore use the negative equivalent width measurements with uncertainties in our analysis, as opposed to truncating the values to zero or limiting our sample based on S/N, to retain the most information from the data.  
 
\ We examine the validity of this model fitting approach by running Monte Carlo simulations to determine if we could recover our best-fit model parameters with the inclusion of negative column densities in our sample.  For each line of sight, we assume the column density is normally distributed around its best-fit model value with $\sigma$ defined by the measured uncertainty.  Assuming these underlying normal distributions, we deviate the column densities from their best-fit model values and find new best-fit parameters for the deviated data.  Figure 6 shows histograms of best-fit parameters $\beta$ and $n_or_c^{3\beta}$ from Equation (6) and the number of negative column densities in each simulation for 20,000 Monte Carlo simulations.  Due to the large uncertainties for several of our column density measurements, our simulations consistently deviate column densities to negative values (median value of 7).  The median values of $\beta$ and $n_or_c^{3\beta}$ are 0.71 and 0.050 cm$^{-3}$ kpc$^{3\beta}$, which are both consistent with our input values of 0.71 and 0.048 cm$^{-3}$ kpc$^{3\beta}$ for $\beta$ and $n_or_c^{3\beta}$, respectively.  Furthermore, the distributions of $\beta$ and $n_or_c^{3\beta}$ are consistent with the 1$\sigma$ boundaries of their best-fit values (see Table 2).  This implies we reproduce our model fitting results with negative column densities in our sample and thus motivates the inclusion of the negative column density measurements in our sample.

\section{Summary and Discussion}

\subsection{Spatial Distribution of the Gas}

\ There has been recent work on the structure of the Milky Way's hot gas halo utilizing the combination of emission ($\approx n_e^2L$) and absorption ($\approx n_eL$) to constrain the density and size of the halo gas.  While our model is more sophisticated than a uniform density halo, these results serve as a foundation for comparing our density profile results.  \cite{bld07} initially used the same $XMM-Newton$ dataset as this paper without targets 27-29 (see Table 1) and found $n_e$ = 9 $\times$ 10$^{-4}$ cm$^{-3}$, $L$ = 19 kpc.  Alternatively, \cite{gupta12} conducted a similar analysis with eight $Chandra$ targets and found $n_e$ = 2 $\times$ 10$^{-4}$ cm$^{-3}$, $L$ = 72 kpc.  Both of these results are more simplistic than our $\beta$-model and do not extend to the virial radius of the Milky Way.  For comparison, we calculate the $\chi^2$ for these models with our expanded dataset, while also including an additional uncertainty of 7.2 m$\AA$, and find $\chi^2$(dof) = 39.5 (27) and 44.7 (27).  Both of these are larger than the $\chi^2$ we find for our best-fit model, $\chi^2$ = 26.0 (26).  Although the halo size is not a parameter in our model, our best-fit model is a statistical improvement over the uniform density models that have previously characterized the halo gas.

\ The best-fit parameters for our density model are also comparable to observations of hot gas halos around other galaxies where structural analysis has been possible.  When comparing the Milky Way's hot halo to that of other galaxies, the core radius and $\beta$ parameters are the most applicable.  The core density acts as a normalization of the overall profile and is physically not as significant as the other two parameters.  \cite{opc03} conducted a study of early-type galaxies and extrapolated hot gas density profiles from the observed X-ray surface brightness profiles.  Their model fitting procedure resulted in $\beta$ values between 0.4 - 0.8 and core radii of less than 1 kpc for majority of their sample, both consistent with our best-fit parameters.  There has also been more detailed work on the individual early-type galaxy NGC 1600.  \cite{sivakoff04} analyzed the diffuse gas around NGC 1600 and found a two-component model to fit the surface brightness profile with $r_{c,in}$ = 4.2 kpc, $\beta_{in}$ = 1.18, $r_{c,out}$ = 7.3 kpc, $\beta_{out}$ = 0.36.  The fitting results of NGC 1600 are odd in that a double $\beta$ profile is not typically required to fit the X-ray surface brightness profiles around galaxies, particularly with $\beta$ values that are inconsistent with analyses of other galaxies.  This may be attributed to NGC 1600 residing in a group environment rather than in isolation.  Detailed structure analysis of hot gas halos around late-type galaxies has been limited to only the most massive spirals.  \cite{ab11} analyzed the hot gas halo around NGC 1961 ($M_{vir}$ = 2.1 $\times$ 10$^{13}$ $M_{\odot}$) and found $\beta$ = 0.47 and $r_c$ = 1.00 kpc.  Similarly, \cite{dai11} have analyzed the hot gas halo of UGC 12591 ($M_{vir}$ = 3.5 $\times$ 10$^{13}$ $M_{\odot}$) and found $\beta$ = 0.48 and $r_c$ = 3.04 kpc.  Our measured $\beta$ of 0.71 is steeper than observations of these massive spirals and the measured core radii for the most massive spirals are larger than the core radius determined for the Milky Way.  The discrepancy between these parameters may be explained by the larger stellar disks and dark matter halos associated with NGC 1961 and UGC 12591 compared to the Milky Way.  The comparison between the core radii and $\beta$ parameters is also limited due to the weak constraint we have on the Milky Way's core radius and the degeneracy between the core radius and core density. 

\ The best-fit parameters from our model fitting procedure result in a density profile that is consistent with previously established constraints.  The density model with best-fit parameters can be seen in Figure 7.  Based on analyzing the pulsar dispersion measure toward the LMC, the average electron density between the Sun and the LMC must be $\langle n_e\rangle \leq 5 \times 10^{-4}$ cm$^{-3}$ \citep{ab10}.  For our density profile, the best-fit parameters result in an average electron density of $\langle n_e\rangle = 1.2 \times 10^{-4} $ cm$^{-3}$ between the Sun and the LMC assuming a solar metallicity.  This is well below the upper limit established by the pulsar dispersion measure.  The combination of having an upper limit to the average electron density between the Sun and the LMC and the fact that the total electron column density scales with metallicity as $\propto$ $Z^{-1}$ (Equation (1)) allows us to place a lower limit on the metallicity of the gas toward the LMC.  We find a minimum metallicity of $Z$ $\gtrsim$ 0.2 $Z_{\odot}$ in order to satisfy the pulsar dispersion measure toward the LMC.  This lower limit is consistent with the metallicities of some high-velocity clouds (HVCs), particularly Complex C, and the Magellanic Stream \citep{gibson00, woerwakk04, fox05}, implying the halo gas metallicity may be predominantly sub-solar.  

\ We also examine the possibility of this hot gas extending out to the Milky Way's virial radius, which would affect satellites of the Milky Way.  There have been numerous studies investigating ram-pressure stripping of dwarf spheroidal galaxies orbiting the Milky Way, which implies the presence of a hot corona associated with the Milky Way out to $\sim$200 kpc \citep{br00,gp09}.  \cite{br00} found that dwarf spheroidal galaxies orbiting the Milky Way will effectively lose their cold gas due to ram-pressure stripping for an ambient halo with density of $n$ $\geq$ 2.4 $\times$ 10$^{-5}$ cm$^{-3}$ out to the virial radius of the Milky Way.  \cite{gp09} found a considerably larger value, 2.5 $\times$ 10$^{-4}$ cm$^{-3}$, out to similar distances.  Our best-fit model results in densities that are too low to satisfy either of these constraints.  In order to account for this we add an ambient density to our model of $n$ = 1 $\times$ 10$^{-5}$ cm$^{-3}$ out to 200 kpc.  We choose a medium consistent with \cite{br00} as opposed to \cite{gp09} because an additional ambient medium as large as 2.5 $\times$ 10$^{-4}$ cm$^{-3}$ violates the observational constraint of the emission measure out of the Galactic plane, which will be discussed in greater detail in Section 4.2.  We note that the addition of this density component does not change the best-fit parameters derived in our model fitting because the low ambient density does not contribute a significant fraction to the observed column densities.  Thus, we can add this ambient medium to our density profile without jeopardizing the validity of our best-fit parameters.

\ We compare the scale height of our density model to the scale heights of different ions that represent different temperature phases of the ISM.  The scale height is defined as

	\begin{equation}
	 h = \left|\frac{n}{dn/dz}\right| = \frac{r_c^2}{3\beta z} \left( 1 + \frac{R^2 + z^2}{r_c^2} \right)
	\end{equation}

\noindent When evaluating the scale height for our density model, we note that the function for the scale height is dependent on both $R$ and $z$ for our spherical model.  We calculate the scale height for our profile at several different $R$ and $z$ distances (Table 3).  The applicable comparison is for $R$ = 8.5 kpc values since previous studies of other ions are inherently observed at the solar circle.  The Li-like ions with peak abundances at temperatures 3, 2 and 1 $\times$ 10$^{5}$ K are O {\scriptsize{VI}}, N {\scriptsize{V}} and C {\scriptsize{IV}} \citep{sd93}.  The scale heights of these ions have been measured to be $h$(O {\scriptsize{VI}}, \textit{b} $\textgreater$ 0) = 4.6, $h$(O {\scriptsize{VI}}, \textit{b} \textless 0) = 3.2, $h$(N {\scriptsize{V}}) = 3.9, and $h$(C {\scriptsize{IV}}) = 4.4 kpc \citep{sav97, bowen08}.  The scale height(s) we determine for O {\scriptsize{VII}} gas at $T$ $\sim 10^6$ K are larger than these cooler ions by about an order of magnitude.  This is expected due to the difference in temperature between the ions being approximately an order of magnitude. Thus, the distribution of our density model is more extended than the 10$^5$ K gas, indicating they are not cospatial \citep{williams05}.  

\subsection{Density and Mass Considerations}

\ The primary goal for determining the density profile of the Milky Way's hot gas halo is to determine the amount of mass it contains.  The best-fit parameters and size of the halo determine whether there is enough mass contained in this temperature gas to account for some or all of the missing baryons in the Milky Way.  The mass profile corresponding to the best-fit density profile can be seen in Figure 8 assuming the gas has a solar metallicity.  We consider the mass contained within 18 kpc and 200 kpc as limits on the minimum and maximum mass of the halo.  The minimum halo size is based on statistical arguments presented in Section 3.2 while we assume the halo extends to the Milky Way's virial radius for a maximum halo size.  Given our best-fit parameters, we find $M$(18 kpc) = 2.2 $^{+6.7}_{-1.3}$ $\times$ 10$^8$ $M_\sun$ and $M$(200 kpc) = 1.2 $^{+1.7}_{-0.2}$ $\times$ 10$^{10}$ $M_\sun$.  The mass contribution from the additional ambient medium discussed in Section 4.1 is substantial at 200 kpc ($\approx$ 7 $\times$ 10$^9$ $M_\sun$), but generally consistent with the error bars on our mass estimates.  The known baryonic mass in the Milky Way (stars + cold gas) is approximately 5 $\times$ 10$^{10}$ $M_\sun$ \citep{bt08}.  This implies stellar + cold gas to hot gas mass fractions of 230 (18 kpc) and 4 (200 kpc).  We then compare the hot gas mass to the virial mass of the Milky Way ($\sim2$ $\times$ 10$^{12}$ $M_\sun$) and define the baryon fraction as $f_b = M_b / M_{tot}$.  The resulting baryon fractions are $f_b$ (18 kpc) = 0.02 $^{+0.01}_{-0.01}$ and $f_b$ (200 kpc) = 0.03 $^{+0.01}_{-0.01}$, both of which are much smaller than the value obtained from the \textit{Wilkinson Microwave Anisotropy Probe} five-year data $f_b$ = 0.171 $\pm$ 0.006 \citep{dunkley09}. 

\ There are several uncertainties in our analysis that can significantly change our mass estimate and corresponding baryon fraction.  The virial mass of the Milky Way has been estimated to be between 1.0 - 2.4 $\times$ 10$^{12}$ $M_\sun$ \citep{bk13}.  If we consider the virial mass of the Milky Way to be 1 $\times$ 10$^{12}$ $M_\sun$ as opposed to 2 $\times$ 10$^{12}$ $M_\sun$, the missing baryon mass is 1.5 $\times$ 10$^{11}$ $M_\sun$ as opposed to 3.6 $\times$ 10$^{11}$ $M_\sun$.  This also changes the virial radius of the Milky Way by about a factor of $\sim$0.8 ($r_{vir}$ $\propto$ $M_{vir}^{1/3}$).  For a virial mass of 1 $\times$ 10$^{12}$ $M_\sun$ and virial radius of 160 kpc, our best-fit model halo results in a hot gas mass and baryon fraction of $M$(160 kpc) = 6.5 $^{+13.2}_{-1.3}$ $\times$ 10$^{9}$ $M_\sun$ and $f_b$ (160 kpc) = 0.05 $^{+0.02}_{-0.00}$.  In this case, the halo gas bound to the Milky Way accounts for 5\% - 15\% of the missing baryons.

\ One possibility that increases the baryon fraction is halo gas extending beyond the Milky Way's virial radius, implying the gas is not bound to the Milky Way.  Given our best-fit parameters and the range of virial masses discussed above, the halo would need to be between 400 - 600 kpc ($\sim$3$r_{vir}$) to account for the missing baryons.  We are unable to rule out a halo this large since our results are insensitive to the low density gas that potentially exists at this radius.  Other studies have explored the possibility of non-local O {\scriptsize{VII}} absorption by examining galaxies who have impact parameters within 2-3 virial radii of a given AGN line of sight \citep{fang06, ab10}.  These nondetections of halo gas result in upper limits on the column densities of halo gas beyond the virial radii of other galaxies.

\ The metallicity of the halo gas also can potentially increase our mass estimates.  We initially assumed a solar gas metallicity in our conversion from O {\scriptsize{VII}} to electron column density. However, we note $N_e$ $\propto$ $Z^{-1}$ (see Equation (4)), implying all of our inferred electron columns will increase if the metallicity is sub-solar.  This effectively changes the normalization of our profile and results in $M$ $\propto$ $Z^{-1}$ for a given halo size.  If we consider a halo gas metallicity of $Z$ = 0.3 $Z_{\odot}$ (within the lower limit established by the pulsar dispersion measure toward the LMC), $M_{vir}$ = 1 $\times$ 10$^{12}$ $M_\sun$ and a halo extending to the virial radius, our mass estimate and baryon fraction become $M$(160 kpc) = 2.2 $^{+4.4}_{-0.5}$ $\times$ 10$^{10}$ $M_\sun$ and $f_b$ (160 kpc) = 0.07 $^{+0.03}_{-0.01}$.  The upper 1$\sigma$ limit on this mass estimate adds a considerable amount of mass to the Milky Way, but only accounts for $\sim$50\% of the missing baryons. 

\ Our mass estimates are comparable to observations of the Milky Way's hot gas halo and simulations of the circumgalactic medium (CGM) around galaxies similar to the Milky Way.  Although previous observations of the Milky Way's hot gas halo have relied upon uniform density approximations (see Section 4.1), the derived masses are consistent with our model parameters with assumptions regarding the gas metallicity.  The model found by \cite{bld07} resulted in a halo gas mass of 4 $\times$ 10$^{8}$ $M_\sun$ for a halo size of 20 kpc, which is consistent with our 1$\sigma$ uncertainties for the enclosed mass at that radius.  Alternatively, \cite{gupta12} found a lower limit on the halo gas mass of $\textgreater$ 6.1 $\times$ 10$^{10}$ $M_\sun$ for $L$ $\textgreater$ 139 kpc assuming the gas metallicity is 0.3 $Z_{\odot}$.  Our halo model predicts a mass between 1.2 - 5.2 $\times$ 10$^{10}$ $M_\sun$ for that same radius and metallicity.  Our mass estimates are also in agreement with simulations of the CGM around Milky Way-sized galaxies if we assume a gas metallicity of $\sim$0.3 $Z_{\odot}$.  Hydrodynamical simulations by \cite{feldmann12} predict CGM densities of $\sim$10$^{-4}$ cm$^{-3}$ out to $\sim$100 kpc, resulting in mass estimates of [0.2, 1.0, 3.5] $\times$ 10$^{10}$ $M_\sun$ at $r$ = [50, 100, 200] kpc.  These estimates are within our 1$\sigma$ uncertainties at each radius for a gas metallicity of $\sim$0.3 $Z_{\odot}$, indicating the halo gas mass is likely comparable to the observed stellar + cold gas mass previously observed for the Milky Way.

\ The enclosed hot gas mass near the disk of the Milky Way is comparable to the observed mass in ionized HVCs.  The total mass of ionized HVCs within 5-15 kpc of the Sun is $M$ $\approx$ 1.1 $\times$ 10$^8(d/12$ kpc)$^2(f_c/0.5)(Z/0.2Z_{\odot})^{-1}$ $M_{\odot}$, where $f_c$ is the covering fraction \citep{lehner12}.  For typical ionized HVC parameters, this mass estimate is approximately equal to the suggested hot gas mass enclosed within 10 kpc of the Galactic center (see Figure 8).  One possibility to explain the similarity between these masses is that the ionized HVCs could have cooled out of the hot halo and are accreting on the disk of the Milky Way.  However, the origin of HVCs is still poorly understood and likely a combination of several sources.  The consistency between the mass estimates offers one possible formation mechanism.  

\ The Milky Way's hot gas halo has been observed in X-ray emission by several groups and our density profile must be consistent with these observational constraints.  We note that because the density profile of the gas falls off faster than $n$ $\propto$ $r^{-1}$ the column density is dominated by gas closest to the Galactic center.  This effect is more drastic when we examine the X-ray emission of the halo due to the emission measure scaling as $n^{2}$.  We define the emission measure as

	\begin{equation}
	 EM = \int_0^d n_en_p ds
	\end{equation}

\noindent where we note that to be consistent with Equations (1)-(3), $d$ is the line of sight distance and $n_e$ and $n_p$ are functions of both $b$ and $l$.

\ One constraint we must address is the X-ray emission measure toward the Galactic center determined by \cite{snow97} using the \textit{ROSAT} all-sky survey.  The observed count rate toward the Galactic center is $\sim$150 $\times$ 10$^{-6}$ counts s$^{-1}$ arcmin$^{-2}$, which includes extinction in the Galactic plane.  However, by assuming an absorbing column of 4.4 $\times$ 10$^{21}$ H I cm$^{-2}$, they extrapolate a peak count rate of $\sim$900 $\times$ 10$^{-6}$ counts s$^{-1}$ arcmin$^{-2}$ for the 3/4 keV band.  Using their conversion between count rate and emission measure, which is sensitive to temperature, we determine that the extrapolated count rate corresponds to an emission measure of 0.45 cm$^{-6}$ pc.  In order to reproduce this emission measure, we need to consider an inner radius where the density is constant for 0 $\leqslant$ $r$ $\leqslant$ $r_{in}$.  We find that our density model must be constant at $n_{in}$ = 8.8 $\times$ 10$^{-3}$ cm$^{-3}$ out to an inner radius of $r_{in}$ = 2.2 kpc to reproduce the observed emission measure toward the Galactic center.  We note that this inner radius does not affect the other parts of our analysis.  In particular, the mass estimate is not affected by this due to the small volume associated with this region.  Also, we note that this inner radius is larger than the core radius of our profile.  This is not a major concern since $r_c$ is not well constrained and we still constrain the extended regions of the profile reasonably well.  

\ The other emission measure constraint of interest is the emission measure out of the Galactic plane.  \cite{mccammon02} analyzed a 1 sr region of sky at \textit{l} = 90$^{\circ}$, \textit{b} = +60$^{\circ}$ using a quantum calorimeter sounding rocket.  Their sensitivity and spectral resolution allowed them to model the soft X-ray diffuse background into an absorbed thermal component with EM = 0.0037 cm$^{-6}$ pc and an unabsorbed thermal component with EM = 0.0088 cm$^{-6}$ pc.  Given our best-fit halo model, the predicted absorbed emission measure is 0.0017 ($Z/Z_{\odot}$) cm$^{-6}$ pc for $r_{halo}$ = 18 kpc and 0.0018($Z/Z_{\odot}$) cm$^{-6}$ pc for $r_{halo}$ = 200 kpc. This implies that the emission is dominated by the gas within $\approx$ 20 kpc of the Galactic center.  The emission measure produced by our best-fit halo model underproduces the observed emission measure near \textit{l} = 90$^{\circ}$, \textit{b} = +60$^{\circ}$ regardless of the halo size we consider and for solar metallicity.  However, the 1$\sigma$ upper limit on our emission measure along this line of sight is 0.0122 ($Z/Z_{\odot}$) cm$^{-6}$ pc, implying our emission measure estimate is consistent with the observed value at the 1$\sigma$ level.  We also do not consider a separate temperature source in our calculation that could add another component to the observed emission measure, which would also explain the initial discrepancy.

\ The addition of the AGN ESO 141-G055 to our target list allows us to discuss our halo model in the context of recent observations by the \textit{Fermi Gamma-ray Space Telescope}, which revealed two large gamma-ray emitting bubbles above and below the galactic plane \citep{suslatfink10}.  These \textit{Fermi} bubbles are aligned with features seen in the \textit{ROSAT} soft X-ray maps and are believed to be interacting with the Galactic halo has as they expand away from the Galactic plane.  Although these \textit{Fermi} bubbles are considerably hotter than the Galactic halo gas, they will still contribute free electrons to lines of sight toward the Galactic center.  For this comparison, we can use the dispersion measure toward the Galactic center, which is sensitive to the total electron density along a given line of sight.  \cite{taycord93} showed that the dispersion measure toward the Galactic center is 650 - 800 cm$^{-3}$ pc, which is thought to be primarily due to free electrons from gas in the 10$^3$ - 10$^4$ K range (the warm ionized medium) \citep{gaensler08}.  The contribution from our Galactic halo model toward the galactic center is only DM = 72 cm$^{-3}$ pc, which is negligible compared to the model expectations.  The contribution from these \textit{Fermi} bubbles also appears to be negligible and even less than that of our Galactic halo model.  \cite{guomathews11} and \cite{yang12} modeled the \textit{Fermi} bubbles, assuming an ambient medium similar to our determined halo model, to recreate the bubbles' observed structure and found an average density in the plane of $\sim$10$^{-3}$ cm$^{-3}$.  This results in a dispersion measure toward the galactic center of DM = 24 cm$^{-3}$ pc.  Thus, both the $Fermi$ bubbles and Galactic halo contribute a small fraction ($\sim$10\%) of the total electrons near the Galactic plane.  

\ The sightline toward ESO 141-G055 has a peculiarity in the ion column densities that are detected which is directly related to the presence of the \textit{Fermi} bubbles.  Most of our present sample shows little or no O {\scriptsize{VIII}} absorption, which allows us to constrain the temperature of the Galactic halo gas.  However, the line of sight toward ESO 141-G055 suggests an enhancement of O {\scriptsize{VIII}}, with a column density ratio of $N$(O {\scriptsize{VIII}})/$N$(O {\scriptsize{VII}}) = 1.4 $\pm$ 0.5.  \cite{yang12} are able to produce the observed O {\scriptsize{VIII}}/O {\scriptsize{VII}} ratio and find that the shocked region of the bubbles is $\sim$10$^8$ K while the interior is 10$^7$ - 10$^8$ K.  This implies little contribution from O {\scriptsize{VII}} or O {\scriptsize{VIII}} to the total electron column along the line of sight.  While the ion fractions of both ions are small inside the bubbles ($f$ $\ll$ 0.1), the O {\scriptsize{VIII}} ion fraction is at least an order of magnitude higher than the O {\scriptsize{VII}} fraction everywhere inside the bubbles.  This results in the enhancement of O {\scriptsize{VIII}} relative to O {\scriptsize{VII}} for any line of sight that passes through the \textit{Fermi} bubbles.  However, these results do not explain the infrequent detection of O {\scriptsize{VIII}} along most of our other sightlines.  A more detailed analysis of the density and temperature structure of the bubbles is beyond the scope of this work and will be the topic of a future project.

\subsection{Thermal Considerations}

\ The thermal properties of the hot gas halo (mainly the cooling time and radius) can be used as a measure of how large the halo could be if it were stable.  We first adopt an expression for the cooling time \citep{fp06}:

	\begin{equation}
	 \tau(r) = \frac{1.5nkT}{\Lambda(T,Z) n_e(n - n_e)} \approx \frac{1.5kT\times 1.92}{\Lambda(T,Z) n_e\times 0.92}
	\end{equation}

\noindent which assumes a primeval helium abundance.  The cooling time as a function of radius can be seen in Figure 9 for different metallicities. 

One result to note is that the cooling time is less than a Hubble time for a solar metallicity halo out to near the virial radius, implying a need for a continuous heating source if the halo were stable.  This can be explained if the primary source of the halo gas is feedback from the disk in the form of supernovae or AGN, which would enrich and heat the halo.  This allows the cooling time of the halo to be less than the Hubble time since the halo would receive an input of energy, making it stable throughout the Milky Way's lifetime.  Alternatively, if the halo gas is primarily accreted material by the Milky Way, the gas metallicity would be sub-solar.  This implies a cooling radius between 30 - 40 kpc for a $\sim$0.3 $Z_{\odot}$ halo, so the halo at $r$ $\textgreater$ $r_{cool}$ could not have cooled since the formation of the Milky Way.  

\ We compare the cooling time as function of radius to the sound crossing time as a function of radius to determine if the Milky Way's hot gas halo is in hydrostatic equilibrium.  We assume that the halo is isothermal at a temperature of log $T$ = 6.1, which results in a sound speed of $c_s \sim$ 130 km s$^{-1}$.   Figure 9 shows that the sound crossing time is smaller than the cooling for $r$ $\gtrsim$ 1 kpc, implying the halo is in hydrostatic equilibrium.  

\ The cooling time discussed above also has implications for the cooling and accretion rates of the Galactic halo gas on the disk of the Milky Way.  The accretion rate is determined by integrating the mass within a radial shell divided by the cooling time at a given radius and can be seen in Figure 10.  The sensitivity of the result on metallicity has various implications.  For a solar metallicity halo, the accretion rate is similar to modeled accretion rates of similar mass spiral galaxies \citep{fratbin08}.  Figure 10 also indicates that the accretion rate is broadly consistent with the current observed star formation rate (SFR) in the Milky Way of 0.68 - 1.45 $M_{\odot}$ yr$^{-1}$ \citep{rw10}.  This implies that the cooling of the Galactic halo may be a significant source of cold gas that fuels star formation in the disk of the Milky Way.  However, there is also evidence indicating the observed Milky Way SFR can be balanced by stellar mass loss alone.  \cite{lk11} modeled mass loss rates for single-age stellar populations and determined star formation histories for several galaxies using the relationship between SFR and stellar mass in star-forming galaxies.  Their results indicate that in most of their sample, including the Milky Way, mass loss from later stages of stellar evolution can more than compensate for the current observed SFRs.  This indicates that the sub-solar metallicity accretion rate of 0.1 - 0.5 $M_{\odot}$ yr$^{-1}$ from the halo is more likely than a solar metallicity accretion rate if the mass supply rate is to be less than the observed SFR.  If the halo has a solar metallicity ($\sim$1.0 $M_{\odot}$ yr$^{-1}$) and stellar mass loss contributes $\sim$1.5 $M_{\odot}$ yr$^{-1}$ back to the Milky Way disk, then the Milky Way's SFR should increase with time. This opposes the observed cosmic SFR, indicating that the halo is likely not entirely at a solar metallicity \citep{borch06}.  Another possibility that prevents the halo cooling rate from overproducing the observed Milky Way SFR is a heating agent for the halo gas, such as supernovae.  The addition of a heating source increases the cooling time of the halo gas, particularly near the stellar disk, and can significantly reduce the amount of gas cooling out of the halo.  

\ The luminosity of the Galactic halo can be determined from the measured cooling rate discussed above.  The conversion between the 0.5 - 2 keV luminosity and cooling rate is

	\begin{equation}
	L_X (r) = 0.362 \times \dot{M} \frac{1.5 kT}{\mu m_p}
	\end{equation}

\noindent where \textit{\.M} is the cooling rate and 0.362 is the conversion between the bolometric luminosity and the 0.5 - 2 keV band luminosity.  For typical cooling rates seen in Figure 10 and a solar gas metallicity, the corresponding 0.5 - 2 keV band luminosity is $\sim$ 7 $\times$ 10$^{39}$ erg s$^{-1}$.  This is larger than what has been determined from \textit{ROSAT} measurements of the diffuse X-ray background \citep{snow97}, which imply $L_X \sim$ 2 $\times$ 10$^{39}$ erg s$^{-1}$.  The difference is likely due to the uncertainty in the gas metallicity.  A solar metallicity halo should be considered as an upper limit to \textit{\.M} since only part of the halo is expected to be enriched.  In order to match the luminosity determined by \textit{ROSAT}, the metallicity would need to be $\sim$0.3 $Z_{\odot}$.  Thus, our luminosity is broadly consistent with previous estimates of the Milky Way's diffuse X-ray luminosity if the average metallicity is less than solar.  

\ If we assume that this hot halo is volume filling we can compare the pressure of this hot halo with other phases of the ISM.  In particular, we compare our model pressure to pressures associated with HVCs measured by \cite{fox05}, which are at temperatures ranging from $10^4 \sim 10^5$ K.  In their analysis, they average six HVC models to find thermal pressures of \textit{P/k} = [530, 140, 50] cm$^{-3}$ K at distances of [10, 50, 100] kpc.  If we assume a temperature of log $T$ = 6.1, our model results in a range of pressures of \textit{P/k} = [694, 41, 24] cm$^{-3}$ K for $r$ = [10, 50, 100] kpc respectively.  This indicates that our hot gas is close to pressure equilibrium with these observed HVCs.  However, it should be noted that the distances toward these HVCs are not well constrained and the results here are strongly dependent on both the density and temperature of the gas.  We also do not consider the addition of a hotter gas phase in our analysis which would add an additional pressure component.  

\subsection{Final Comments}

\ The goal of this study was to constrain the density profile of the Milky Way's hot gas halo better than previous studies, which have primarily relied upon simple models of the halo structure.  We use \textit{XMM-Newton} RGS archival data of 29 sightlines to analyze O {\scriptsize{VII}} absorption from the halo.  One limitation of our analysis is the inability of the RGS to resolve the observed absorption lines.  This prohibits us from analyzing the true saturation effects in the lines.  From Table 2, accounting for line saturation with a Doppler width of 150 km s$^{-1}$ increases $n_{\circ}$ of our density model but steepens the best-fit $\beta$ compared to the optically thin fitting results.  The parameter $n_{\circ}$ increases since all of the inferred column densities increase if we assume any line saturation.  The parameter $\beta$ also increases since the lines that have the largest equivalent widths, and thus the largest inferred column densities, also have large uncertainties in the curve of growth analysis.  Table 4 shows the most important results of our analysis assuming both optically thin and saturated best-fit parameters from Table 2.  By using saturated column densities, the steeper $\beta$ parameter is more important than the increased normalization relative to the optically thin results in terms of the overall mass estimate.  However, the inferred masses assuming the lower metallicity limits established by the pulsar dispersion measure toward the LMC result in comparable masses for each case.  The emission measure estimates toward \textit{l} = 90$^{\circ}$, \textit{b} = +60$^{\circ}$ differ by a factor of $\sim$2, which is also due to the steeper $\beta$ in the saturated parameter case.  Neither of these estimates overproduce the observed emission measure for a solar metallicity, implying an additional component to the observed emission measure.  By comparing our best-fit results for optically thin and saturated column densities, we find the best-fit parameters change, but the inferred masses from the best-fit profiles are similar.

\ The metallicity of the halo gas has not been thoroughly analyzed for the Milky Way's halo and is crucial for understanding various properties of the halo gas.  Although we initially assumed a solar metallicity halo in our analysis, we recognize that the metallicity of the gas is a large uncertainty in our analysis and is likely sub-solar.  We are able to place a constraint on the metallicity of the gas between the Sun and the LMC based on the observed pulsar dispersion measure \citep{ab10}.  The metallicity of the gas must be $Z$ $\gtrsim$ 0.2 $Z_{\odot}$ to satisfy the pulsar dispersion measure constraint.  This lower limit applies to the average metallicity of the gas between the Sun and the LMC ($r$ $\approx$ 55 kpc) and does not necessarily apply to halo gas beyond the LMC.  Our results for the mass accretion rate and X-ray luminosity of the halo suggest that a halo metallicity of $Z$ = 0.3 $Z_{\odot}$ is more appropriate.  This metallicity is consistent with cosmological simulations \citep{toft02, co06} and observations of both spiral galaxies \citep{rasmussen09, meiring13} and some HVCs \citep{gibson00, woerwakk04, fox05}.  We also ignore a metallicity gradient in our analysis, which is possible from the mixing of ejected gas from the disk of the Milky Way and cooling primordial gas from the formation of the Milky Way.  Both mechanisms are likely contributing to the halo gas, but the metallicity gradient of the halo gas is not well understood.  An analysis of the halo gas metallicity and on the extent there exists a metallicity gradient will be critical in determining several halo gas properties.

\ With the density profile of the halo constrained, we are able to analyze useful properties of the halo and determine how the halo relates to the baryon content of the Milky Way.  We find the mass contained in the halo for our best-fit parameters is between $M$(18 kpc) = 7.3 $^{+22.3}_{-4.3}$ $\times$ 10$^8$ $M_\sun$ and $M$(200 kpc) = 4.0 $^{+5.7}_{-0.7}$ $\times$ 10$^{10}$ $M_\sun$ for an assumed metallicity $Z$ = 0.3 $Z_{\odot}$.  If we assume a lower estimate of the virial mass of the Milky Way (1 $\times$ 10$^{12}$ $M_\sun$) and the gas extending to the virial radius of the Milky Way for a halo that size, the largest baryon fraction we obtain is $f_b$ (160 kpc) = 0.07 $^{+0.03}_{-0.01}$.  This accounts for 10\% - 50\% of the missing baryons required to match the universal baryon fraction of $f_b$ = 0.171.

\ The constraints we place on the Milky Way's hot gas halo are close to the best we are able to accomplish with \textit{Chandra} and \textit{XMM-Newton}.  Improvements can be made on eliminating the degeneracy between the parameters $n_{\circ}$ and $r_c$ with additional Galactic targets near the Galactic center, however this does not affect the global properties of the halo.  There is also work to be done exploring the interaction between the hot gas halo and the \textit{Fermi} bubbles.  The combination of O {\scriptsize{VII}} and O {\scriptsize{VIII}} emission will reveal the temperature structure just outside and inside the bubbles, which will help probe the contribution of thermal and non-thermal electrons inside the bubbles.  

\acknowledgments
The authors would like to acknowledge valuable advice and conversations with Mike Anderson, Jimmy Irwin, Jon Miller, Mateusz Ruszkowski, Steve Snowden and Hsiang-Yi Yang.  Financial support is gratefully acknowledged from NASA under the ADAP program.

\bibliographystyle{apj}


\addtolength{\oddsidemargin}{0in}
\addtolength{\evensidemargin}{0in}
\addtolength{\textwidth}{1.75in}

\begin{deluxetable}{l l r r r r c c c c c c}

\rotate
\tabletypesize{\tiny}
\tablewidth{0pt}
\tablecolumns{12}
\tablecaption{Absorption-Line Measurements}
\tablehead{
  \colhead{Number} &
  \colhead{Name} &
  \colhead{\textit{l}} &
  \colhead{\textit{b}} &
  \colhead{EW} &
  \colhead{Error} &
  \colhead{$N_{OVII,thin}$} &
  \colhead{Error} &
  \colhead{$N_{model,thin}$} &
  \colhead{$N_{OVII,saturated}$} &
  \colhead{Error} &
  \colhead{$N_{model,saturated}$} \\
  \colhead {} &
  \colhead {} &
  \colhead {($^{\circ}$)} &
  \colhead {($^{\circ}$)} &
  \colhead {(m$\AA$)} &
  \colhead {(m$\AA$)} &
  \colhead {\tiny($10^{15}$ cm$^{-2}$)} &
  \colhead {\tiny($10^{15}$ cm$^{-2}$)} &
  \colhead {\tiny($10^{15}$ cm$^{-2}$)} &
  \colhead {\tiny($10^{15}$ cm$^{-2}$)} &
  \colhead {\tiny($10^{15}$ cm$^{-2}$)} &
  \colhead {\tiny($10^{15}$ cm$^{-2}$)} 
}

\startdata
1   & Mrk 421        & 179.83  &  65.03  &  11.8  &  0.8  &  4.12  &  2.53  &  4.62  &  5.36  &  3.61  &  4.54  \\
2   & PKS 2155-304   & 17.73   &  -52.24  &  13.7  &  1.9  &  4.79  &  2.60  &  7.55  &  6.56  &  4.06  &  8.97  \\
3   & 3C 273         & 289.95  &  64.36  &  24.6  &  3.3  &  8.60  &  2.77  &  5.67  &  18.06  &  9.28  &  5.98  \\
4   & MCG-6-30-15    & 313.29  &  27.68  &  32.6  &  6.8  &  11.36  &  3.45  &  7.73  &  42.35  &  27.19  &  9.28  \\
5   & LMC X-3        & 273.57  &  -32.08  &  21.0  &  5.0  &  7.34  &  3.07  &  3.91  &  13.00  &  7.37  &  4.39  \\
6   & 1 H 1426+428   & 77.49   &  64.90  &  11.6  &  4.1  &  4.04  &  2.90  &  5.52  &  5.21  &  3.99  &  5.78  \\
7   & Ark 564        & 92.14   &  -25.34  &  12.3  &  4.6  &  4.29  &  2.98  &  5.25  &  5.63  &  4.23  &  5.38  \\
8   & NGC 4051       & 148.88  &  70.09  &  24.6  &  3.1  &  8.59  &  2.74  &  4.80  &  18.02  &  9.18  &  4.77  \\
9   & NGC 3783       & 287.46  &  22.95  &  24.1  &  7.5  &  8.40  &  3.64  &  6.05  &  17.17  &  10.64  &  6.56  \\
10  & NGC 5548       & 31.96   &  70.50  &  7.0  &  6.8  &  2.43  &  3.45  &  6.07  &  2.79  &  6.06  &  6.59  \\
11  & Ark 120        & 201.69  &  -21.13  &  -6.0  &  5.5  &  -2.08  &  3.16  &  4.16  &  -2.34  &  5.11  &  3.94  \\
12  & PKS 0558-504   & 257.96  &  -28.57  &  21.7  &  7.8  &  7.58  &  3.70  &  4.97  &  13.83  &  8.86  &  5.00  \\
13  & Mrk 766        & 190.68  &  82.27  &  0.1  &  6.8  &  0.05  &  3.45  &  5.06  &  0.07  &  4.27  &  5.12  \\
14  & NGC 4593       & 297.48  &  57.40  &  23.4  &  8.5  &  8.16  &  3.88  &  5.96  &  16.10  &  10.46  &  6.42  \\
15  & 3C 390.3       & 111.44  &  27.07  &  27.4  &  7.3  &  9.56  &  3.57  &  4.75  &  23.65  &  14.46  &  4.71  \\
16  & NGC 7469       & 83.10   &  -45.47  &  1.6  &  8.9  &  0.57  &  4.01  &  5.51  &  0.59  &  5.57  &  5.75  \\
17  & Mrk 509        & 35.97   &  -29.86  &  25.9  &  7.3  &  9.04  &  3.57  &  8.62  &  20.36  &  12.39  &  10.89  \\
18  & 3C 120         & 190.37  &  -27.40  &  13.8  &  9.2  &  4.81  &  4.09  &  4.15  &  6.59  &  5.83  &  3.94  \\
19  & NGC 3516       & 133.24  &  42.40  &  22.0  &  13.4  &  7.66  &  5.30  &  4.52  &  14.12  &  11.38  &  4.40  \\
20  & Ton 1388       & 223.36  &  68.21  &  34.5  &  15.7  &  12.04  &  6.03  &  4.83  &  54.68  &  45.38  &  4.82  \\
21  & 1H 0414+009    & 191.82  &  -33.16  &  -3.1  &  14.8  &  -1.07  &  5.73  &  4.20  &  -1.14  &  10.16  &  4.00  \\
22  & MR 2251-178    & 46.20   &  -61.33  &  39.8  &  19.6  &  13.90  &  7.28  &  6.25  &  119.63  &  108.60  &  6.86  \\
23  & IC 4329a       & 317.50  &  30.92  &  33.8  &  19.3  &  11.78  &  7.20  &  7.93  &  49.41  &  43.14  &  9.65  \\
24  & Fairall 9      & 295.07  &  -57.83  &  31.1  &  16.4  &  10.84  &  6.27  &  5.89  &  35.30  &  29.08  &  6.31  \\
25  & MS 0737.9+7441 & 140.27  &  29.57  &  -13.8  &  20.7  &  -4.83  &  7.65  &  4.34  &  -6.65  &  6.57  &  4.18  \\
26  & 3C 59          & 142.04  &  -30.54  &  60.9  &  19.2  &  21.24  &  7.15  &  4.33  &  2599.11  &  2463.75  &  4.16  \\
27  & ESO 141-G055   & 338.18  &  -26.71  &  21.4  &  5.3  &  7.48  &  3.12  &  10.75  &  13.48  &  7.69  &  15.06  \\
28  & 4U 1820-30     & 2.79    &  -7.91  &  23.9  &  3.6  &  8.36  &  2.81  &  7.98  &  16.94  &  8.79  &  18.54  \\
29  & X1735-444      & 346.05  &  -6.99  &  24.7  &  9.7  &  8.61  &  4.23  &  5.02  &  18.15  &  12.32  &  9.98  
\enddata
\tablecomments{Our sample consists of 26 AGN, two Galactic sources and one LMC source.  The targets are listed in order of decreasing S/N with the exception of 27-29.  These three targets are additions to the sample used by \cite{bld07}.  ESO 141-G055 is an AGN while 4U 1820-30 (located in the globular cluster NGC 6624) and X1735-444 are Galactic X-ray sources.  All targets are used in our model fitting and analysis.  The $thin$ and $saturated$ subscripts refer to column density conversions assuming the lines are optically thin or saturated assuming a constant Doppler width of 150 km s$^{-1}$.  The $model$ subscripts refer to the column densities along each line of sight resulting from the best-fit parameters found in Table 2.}
\end{deluxetable}


\addtolength{\oddsidemargin}{0in}
\addtolength{\evensidemargin}{0in}
\addtolength{\textwidth}{-1.75in}

\begin{deluxetable}{ l c c c c c }

\tablewidth{0pt}
\tablecolumns{6}
\tablecaption{Model Fitting Results}
\tablehead{
  \colhead{Model}   &
  \colhead{$n_o$} &
  \colhead{$r_c$ / $R_c$, $z_c$}   &
  \colhead{$\beta$}   &
  \colhead{$n_or_c^{3\beta}$}  &
  \colhead{$\chi^{2}$ (dof)}  \\
  \colhead{}   &
  \colhead{(cm$^{-3}$)}   &
  \colhead{(kpc)}   &
  \colhead{}   &
  \colhead{(cm$^{-3}$ kpc$^{3\beta}$)}   &
  \colhead{}
}

\startdata
Spherical - optically thin \tablenotemark{a}   & $0.09^{+0.14}_{-0.06}$ & $0.33^{+0.25}_{-0.23}$ & $0.56^{+0.10}_{-0.12}$ &  $0.013^{+0.016}_{-0.010}$ & 31.0 (26) \\
Spherical - saturated      \tablenotemark{b}   & $0.46^{+0.74}_{-0.35}$ & $0.35^{+0.29}_{-0.27}$ & $0.71^{+0.13}_{-0.14}$ &  $0.049^{+0.341}_{-0.047}$ & 26.0 (26) \\
Approximated               \tablenotemark{c}   & --                     &--                      & $0.71^{+0.17}_{-0.20}$ &  $0.048^{+0.085}_{-0.037}$ & 26.0 (27) \\
Flattened                  \tablenotemark{d}   & $0.46^{+0.74}_{-0.35}$ & $0.42^{+0.16}_{-0.10}, 0.26^{+0.13}_{-0.09} $   & $0.71^{+0.13}_{-0.14}$ & -- & 29.4 (25) \\
Truncated sample           \tablenotemark{e}   & $0.48^{+0.71}_{-0.43}$ & $0.33^{+0.25}_{-0.23}$ & $0.70^{+0.14}_{-0.14}$ &  $0.046^{+0.255}_{-0.045}$ & 26.0 (26) \\
S/N cut sample             \tablenotemark{f}   & $0.51^{+0.72}_{-0.46}$ & $0.20^{+0.18}_{-0.17}$ & $0.60^{+0.12}_{-0.13}$ &  $0.028^{+0.124}_{-0.027}$ & 20.0 (20) 

\enddata

\tablenotetext{a}{Results assuming the absorption lines are optically thin and a spherical density profile described by Equation (5).}
\tablenotetext{b}{Results assuming the absorption lines are saturated with a Doppler width of 150 km s$^{-1}$ and a spherical density profile described by Equation (5).}
\tablenotetext{c}{Results assuming an approximated spherical density profile described by Equation (6).}
\tablenotetext{d}{Results assuming a flattened density profile described by Equation (7).}
\tablenotetext{e}{Same model as the spherical - saturated case, but with negative equivalent width measurements truncated at 0.  }
\tablenotetext{f}{Same model as the spherical - saturated case, but while only analyzing observations in our sample with S/N greater than 1.1.  }
\end{deluxetable}

\begin{deluxetable}{c c c c c c c }

\tablewidth{0pt}
\tablecolumns{7}
\tablecaption{Scale Heights}
\tablehead{
  \colhead{}   &
  \colhead{}   &
  \colhead{}   &
  \colhead{}   &
  \colhead{$R$ (kpc)}   &
  \colhead{}   &
  \colhead{}   \\
  \colhead{}   &
  \colhead{}   &
  \colhead{0}   &
  \colhead{1}   &
  \colhead{5}   &
  \colhead{8.5}   &
  \colhead{20}
}

\startdata
            & 1 & 0.5 & 1.0 & 12.2 & 34.4 & 187.8 \\
            & 2 & 1.0 & 1.2 & 6.8  & 17.9 & 94.6 \\
$z$ (kpc)   & 3 & 1.4 & 1.6 & 5.3  & 12.7 & 63.8  \\
            & 4 & 1.9 & 2.0 & 4.8  & 10.3 & 48.7  \\
            & 5 & 2.3 & 2.4 & 4.7  & 9.1  & 39.8
\enddata

\end{deluxetable}


\begin{deluxetable}{lcc}

\tablewidth{0pt}
\tablecolumns{3}
\tablecaption{Saturation Effects}
\tablehead{
 \colhead{}  &
 \colhead{Optically Thin}  &
 \colhead{Saturated}
}

\startdata
Added uncertainty to EW (m{$\AA$})  &  7.5  &  7.2  \\
$M$(18 kpc) ($M_\sun$)              &  3.1 $^{+6.8}_{-1.9}$ $\times$ 10$^8$     &  2.2 $^{+6.7}_{-1.3}$ $\times$ 10$^8$  \\
$M$(200 kpc) ($M_\sun$)             &  2.4 $^{+4.9}_{-0.5}$ $\times$ 10$^{10}$  &  1.2 $^{+1.7}_{-0.2}$ $\times$ 10$^{10}$  \\
$Z_{LMC}$ ($Z_{\odot}$) \tablenotemark{a}             &  0.4  &  0.2  \\
EM$_{Galactic Pole}$($Z/Z_\sun$) cm$^{-6}$ pc \tablenotemark{b} &  0.0038  &  0.0018

\enddata
\tablenotetext{a}{\textnormal{L}ower limit placed on the gas metallicity based on the pulsar dispersion measure toward the LMC. }
\tablenotetext{b}{Model emission measures toward \textit{l} = 90$^{\circ}$, \textit{b} = +60$^{\circ}$.   }
\end{deluxetable}

\begin{figure}[b]

\centering
\plottwo{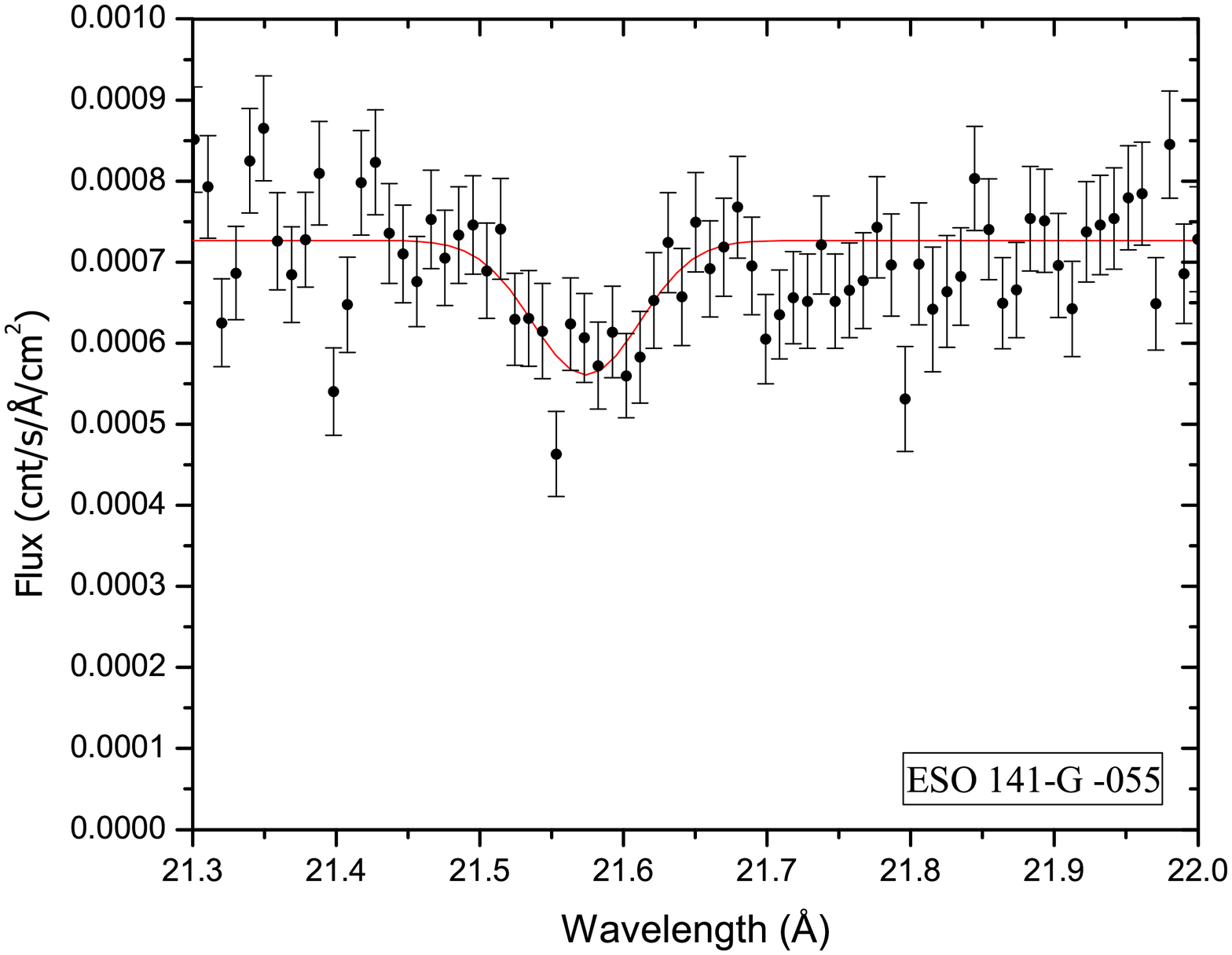}{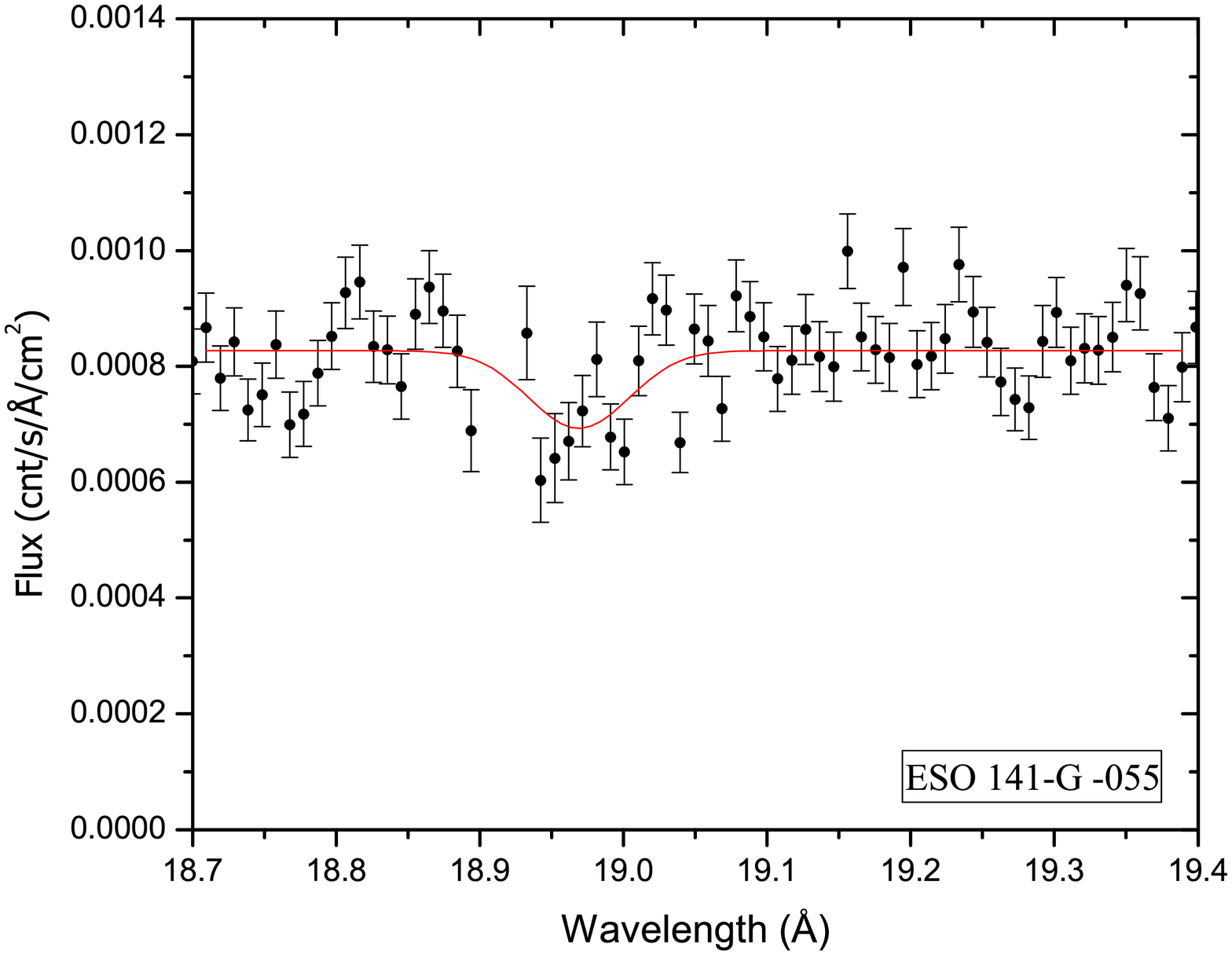}
\plottwo{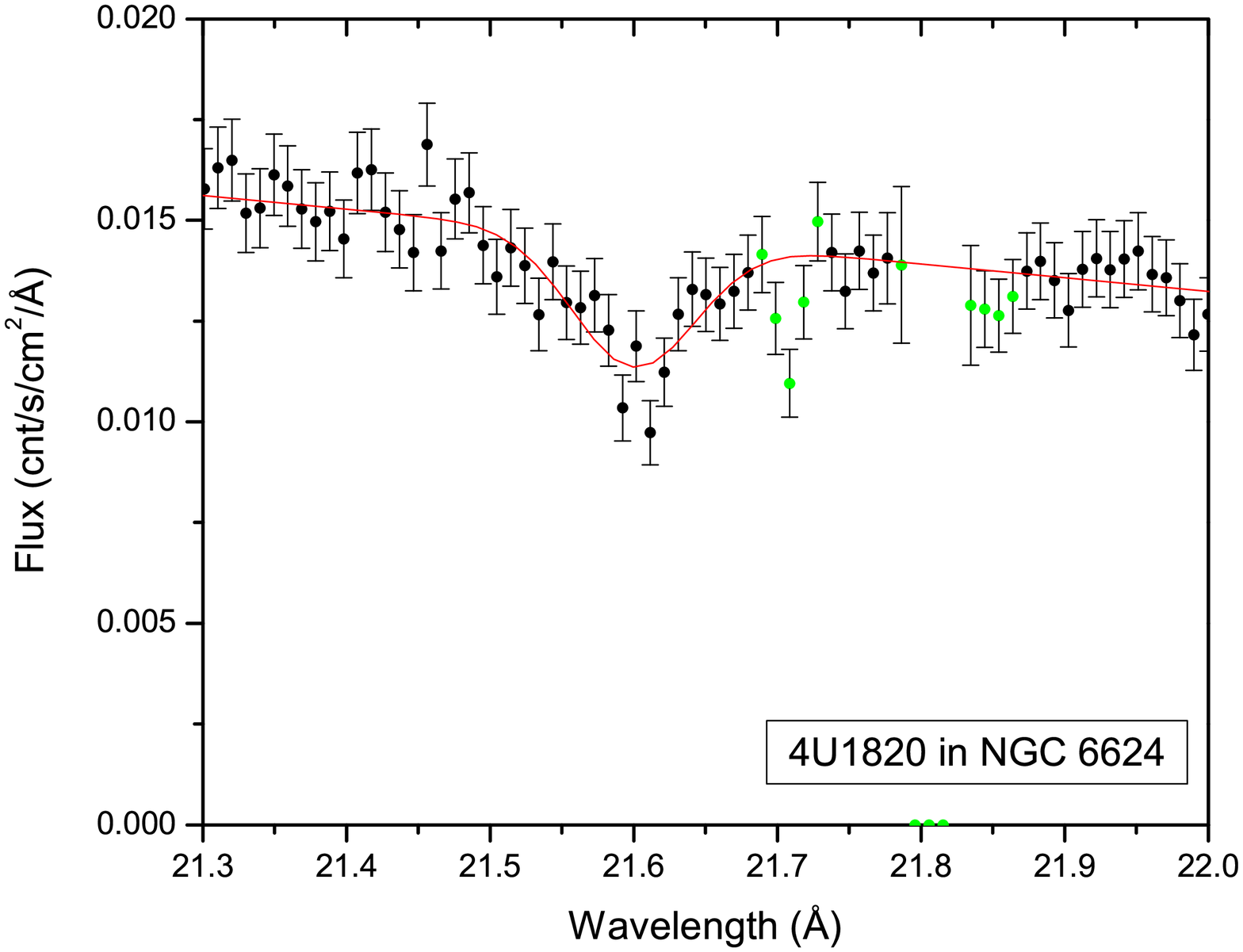}{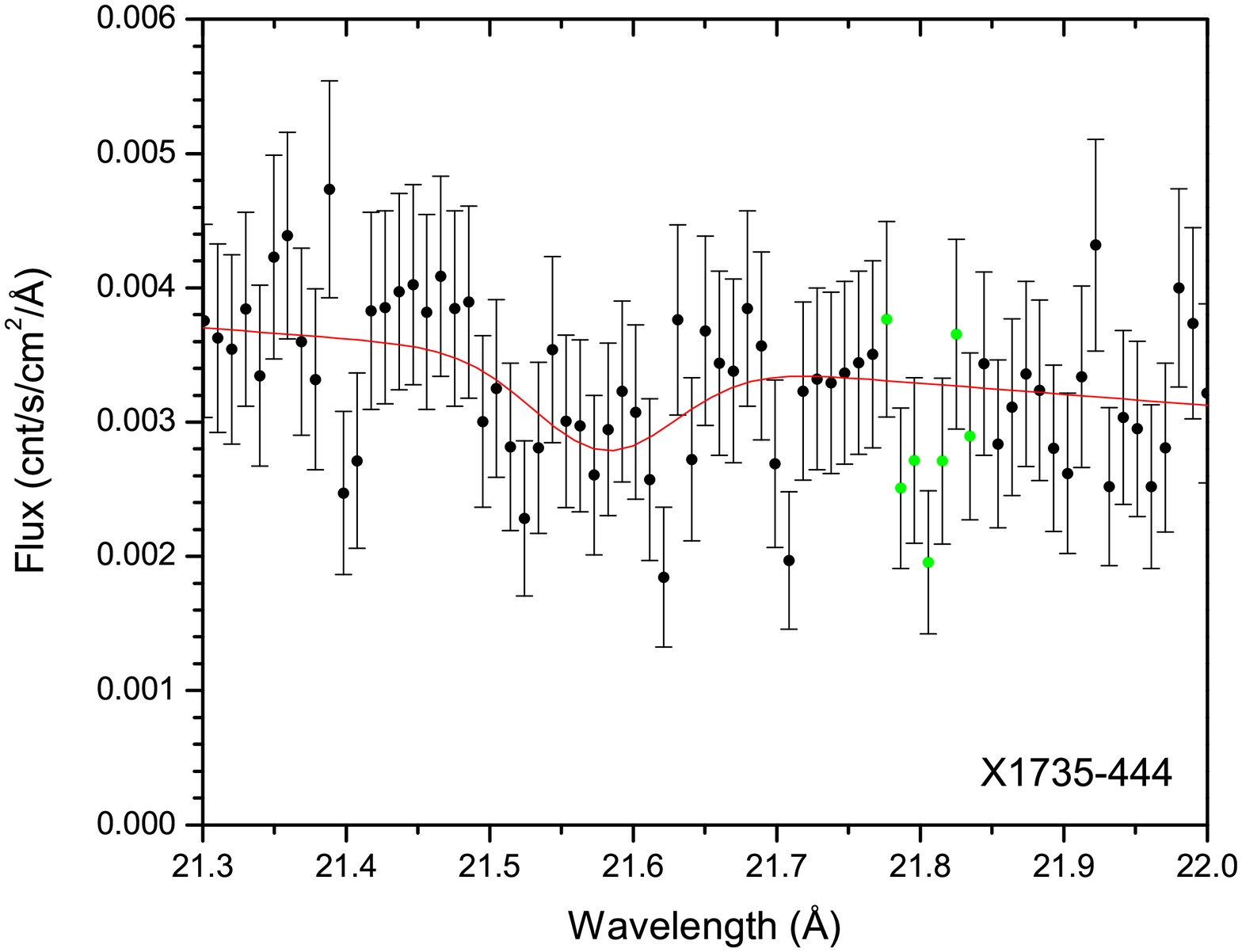}
\caption{\textit{XMM-Newton} flux of our additional targets at 21.60 {$\AA$} (and 18.97 {$\AA$} for ESO 141-G055) to show O {\scriptsize{VII}} and O {\scriptsize{VIII}} absorption.  The continuum and line fitting procedure is the same used by \cite{bld07}.  There are instrumental features in the RGS near 21.82 {$\AA$} and 18.91 {$\AA$} (green points) that are not included in the continuum fitting procedure.  }

\end{figure}


\begin{figure}

\centering
\includegraphics[height =3.4in, width =8.0in, keepaspectratio=true]{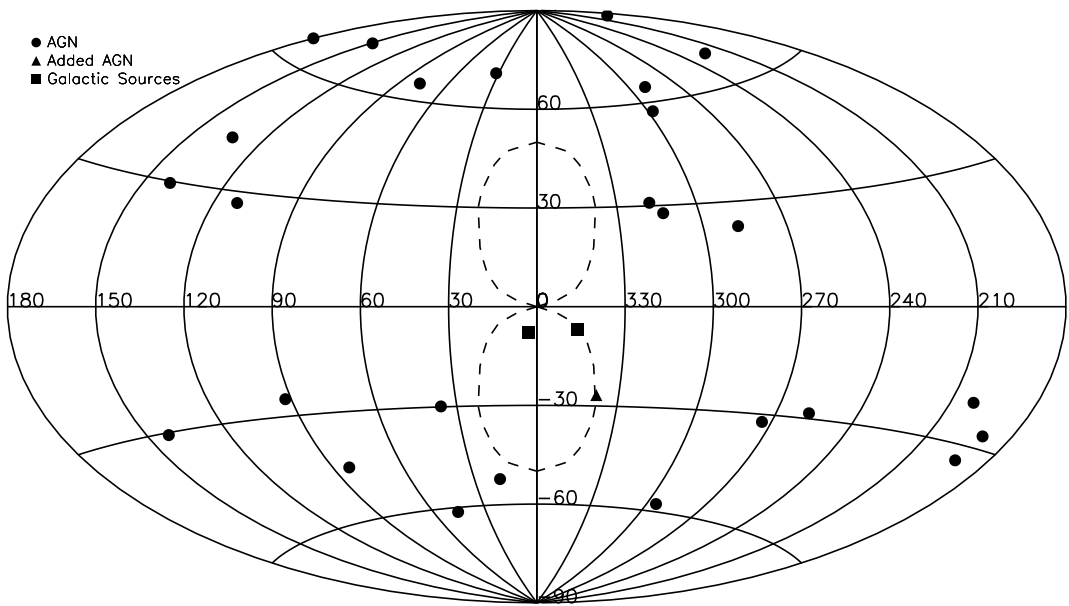}
\caption{Distribution of our X-ray absorbing lines of sight on the sky.  The sample from \cite{bld07} can be seen with solid circles while our additional targets can be seen as squares (Galactic sources) and a triangle (AGN).  The dashed line represents the approximate edges of the north and south \textit{Fermi} bubbles.  The lines of sight of the three added targets pass through the south bubble and allow us to analyze the bubbles' density and temperature structure.  }

\end{figure}


\begin{figure}

\centering
\includegraphics[height =3.4in, width =8.0in, keepaspectratio=true]{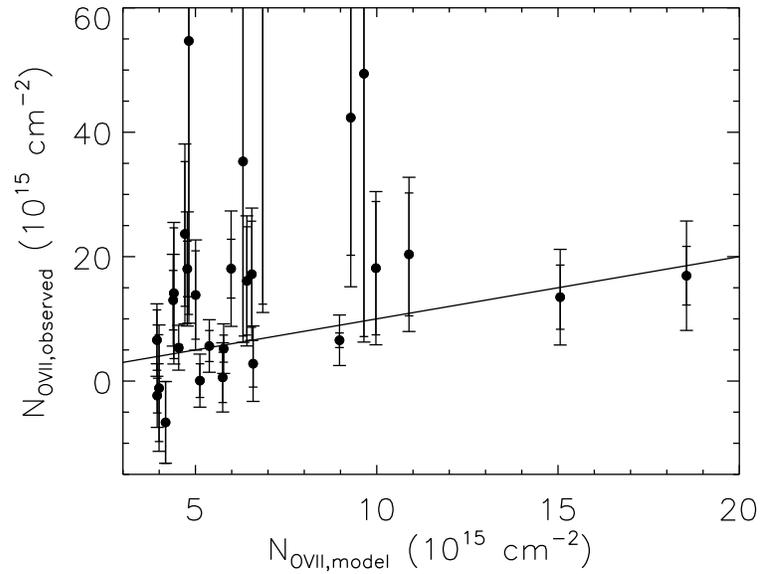}
\caption{Comparison between our observed O {\scriptsize{VII}} column densities and best-fit model column densities in the saturated line case.  The solid line indicates where the observed column density equals the model column density.  The larger errors with larger tick marks for each point represent the initial error with the additional 7.2 m$\AA$ added to each target in order to obtain an acceptable $\chi^2$.  For clarity, the targets MR 2251-178 and 3C 59 are not visible on the plot due to their large observed equivalent widths.  However these lines of sight also have very large uncertainties and are both within 2$\sigma$ of their model column densities.  }

\end{figure}


\begin{figure}

\centering
\plottwo{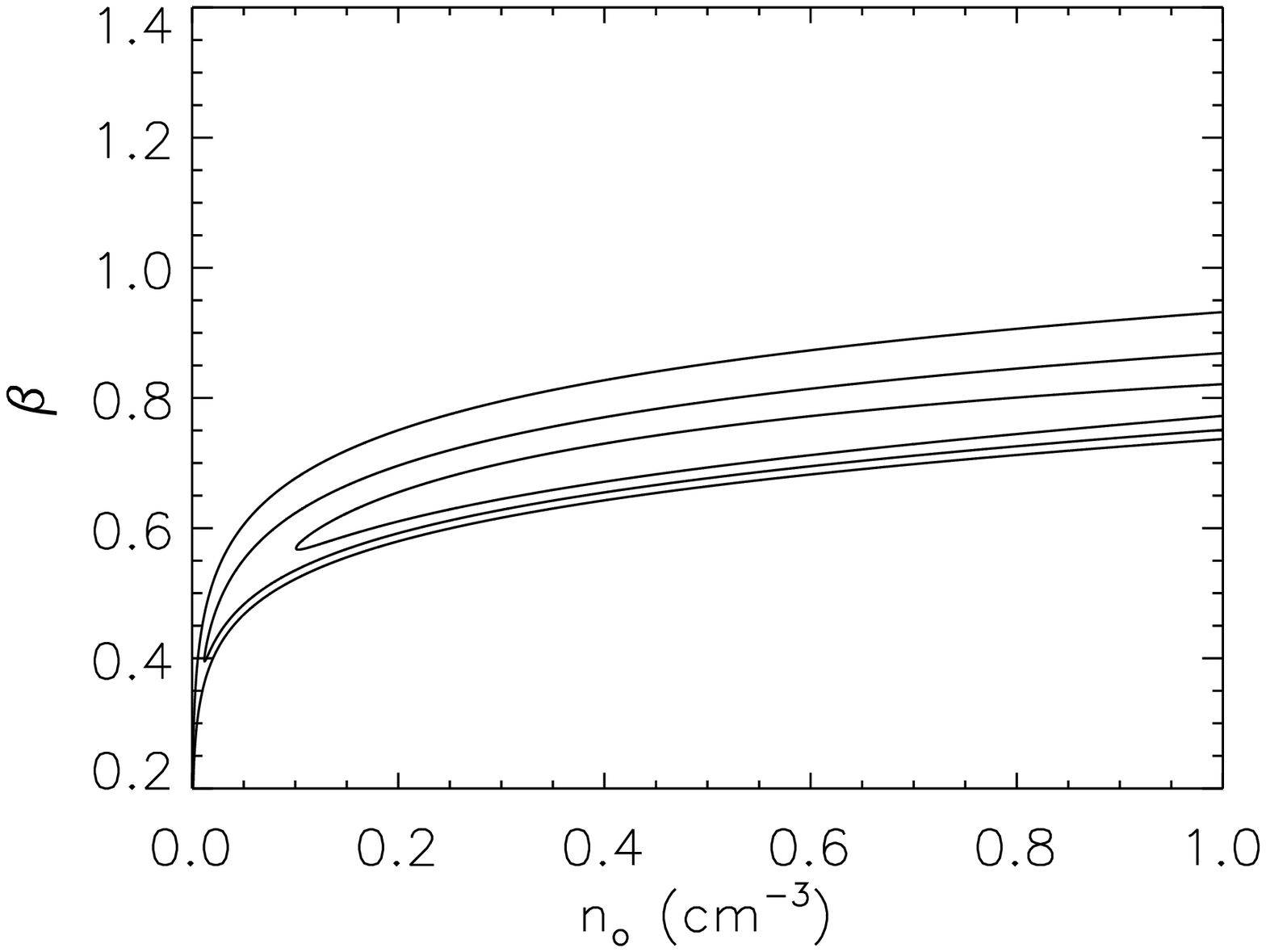}{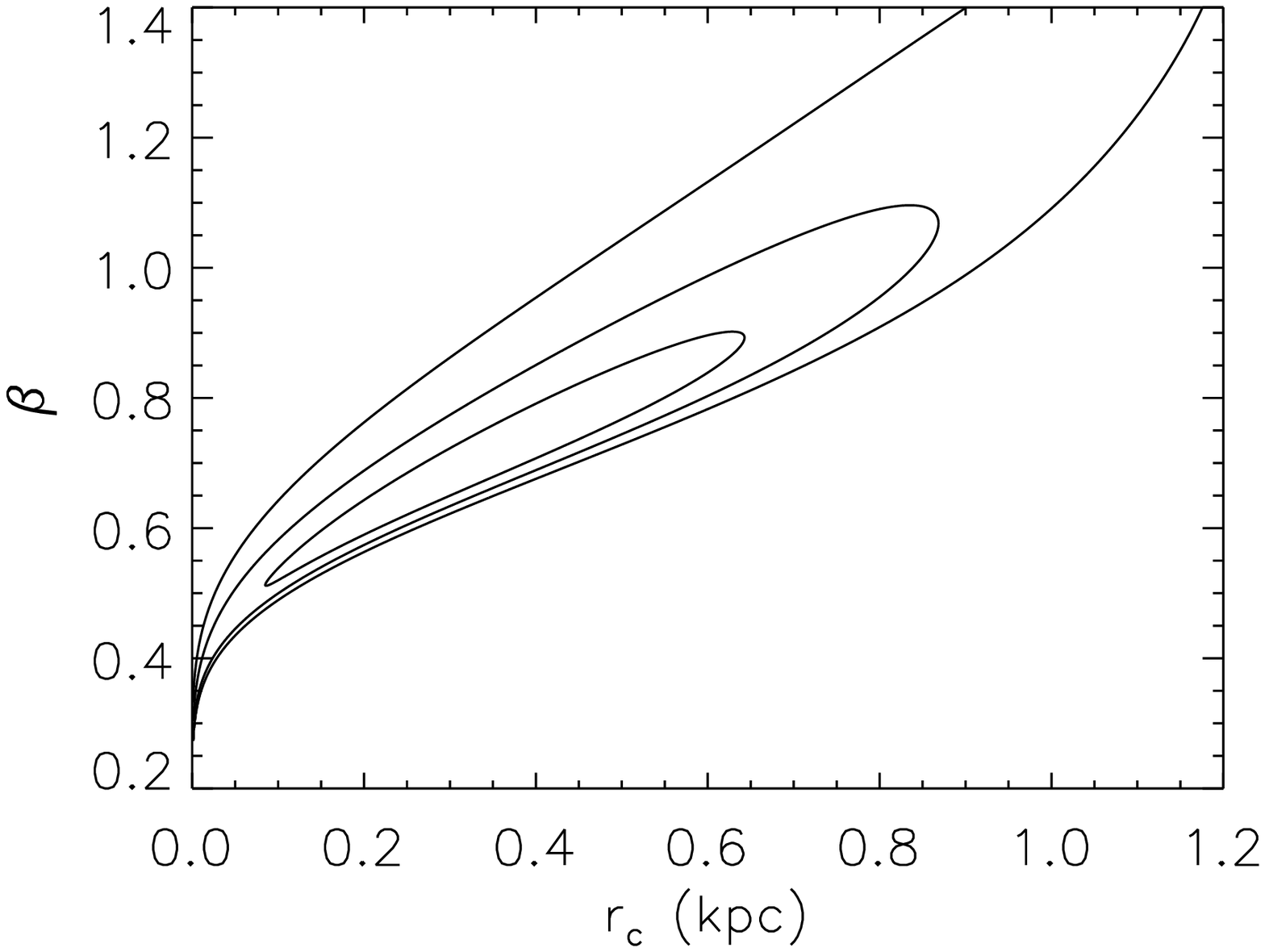}
\plottwo{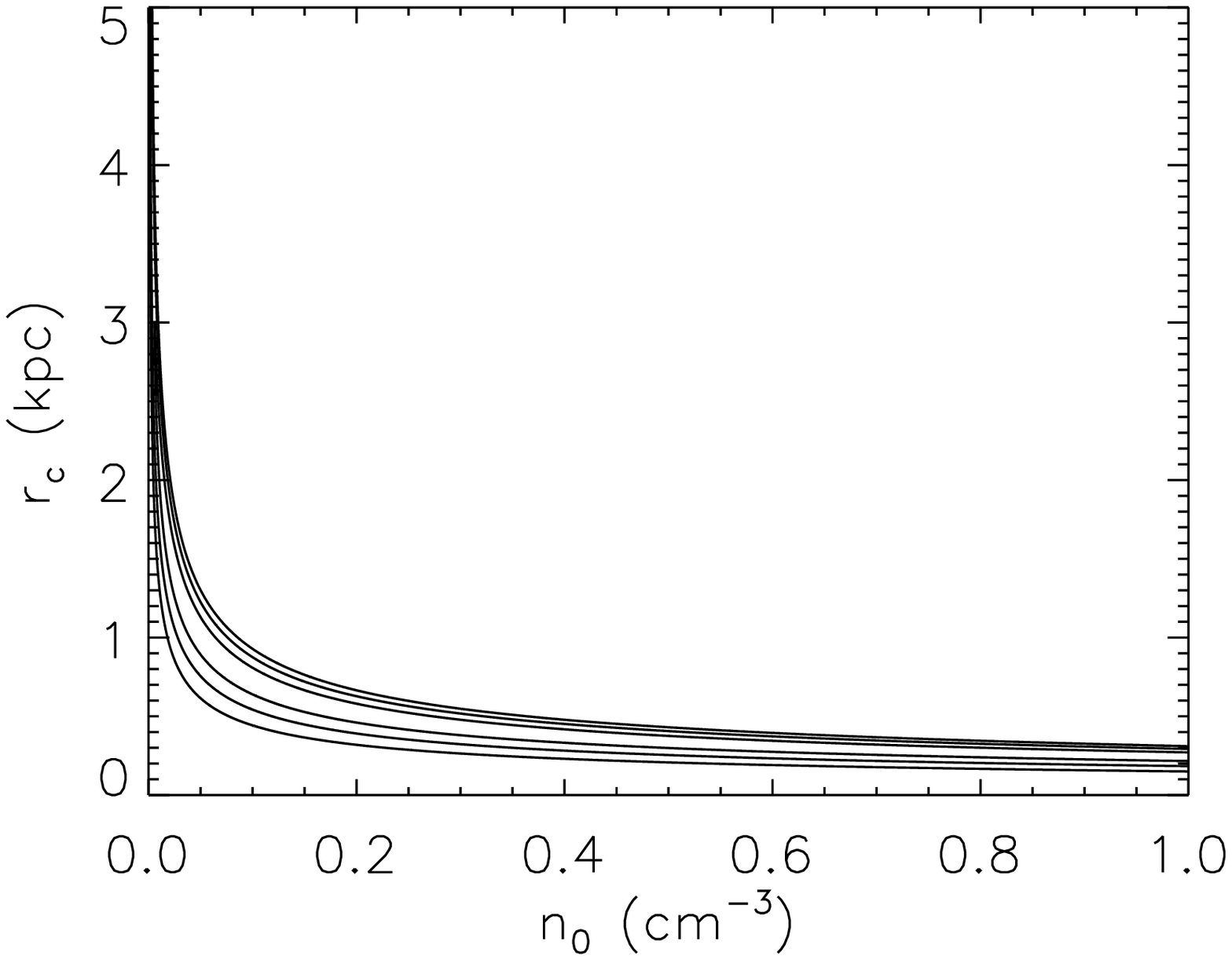}{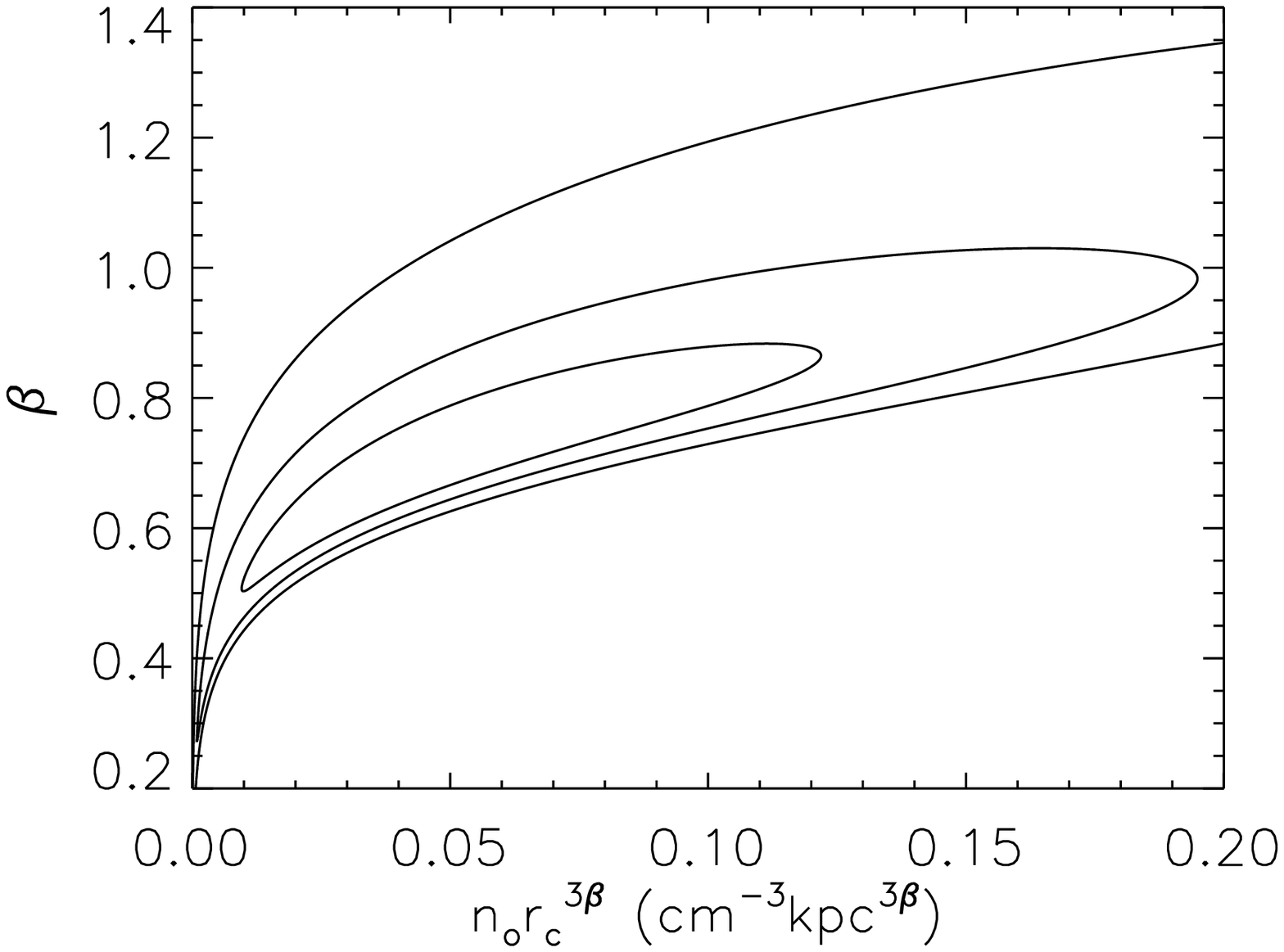}
\caption{1$\sigma$, 2$\sigma$, and 3$\sigma$ contours for $r_c$, $n_o$, $\beta$, and $n_or_c^{3\beta}$ for our spherical model and using saturated column densities assuming $b$ = 150 km s$^{-1}$.  The elongation of the contours in the $r_c$ - $n_o$ plane illustrates the degeneracy discussed in Section 3.2.  The contours constraining $n_or_c^{3\beta}$ are based on the parameters in Equation (2).}

\end{figure}


\begin{figure}

\centering
\includegraphics[height =3.4in, width =8.0in, keepaspectratio=true]{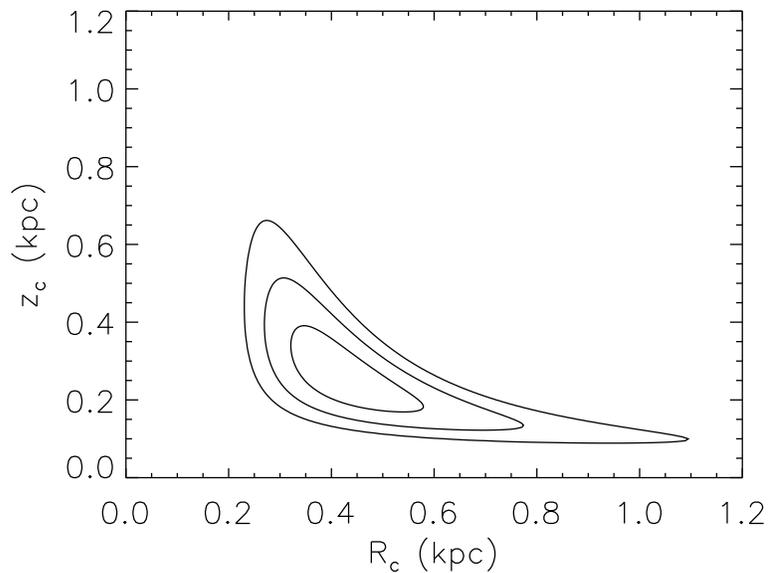}
\caption {1$\sigma$, 2$\sigma$, and 3$\sigma$ contours for $R_c$ and $z_c$ for our flattened model.  $R_c$ corresponds to the core radius in the disk of the Milky Way and $z_c$ corresponds to the core radius out of the plane of the Milky Way.  The shape of the contours indicates that the halo is preferentially aligned with the disk of the Milky Way.  }

\end{figure}


\begin{figure}

\centering
\plottwo{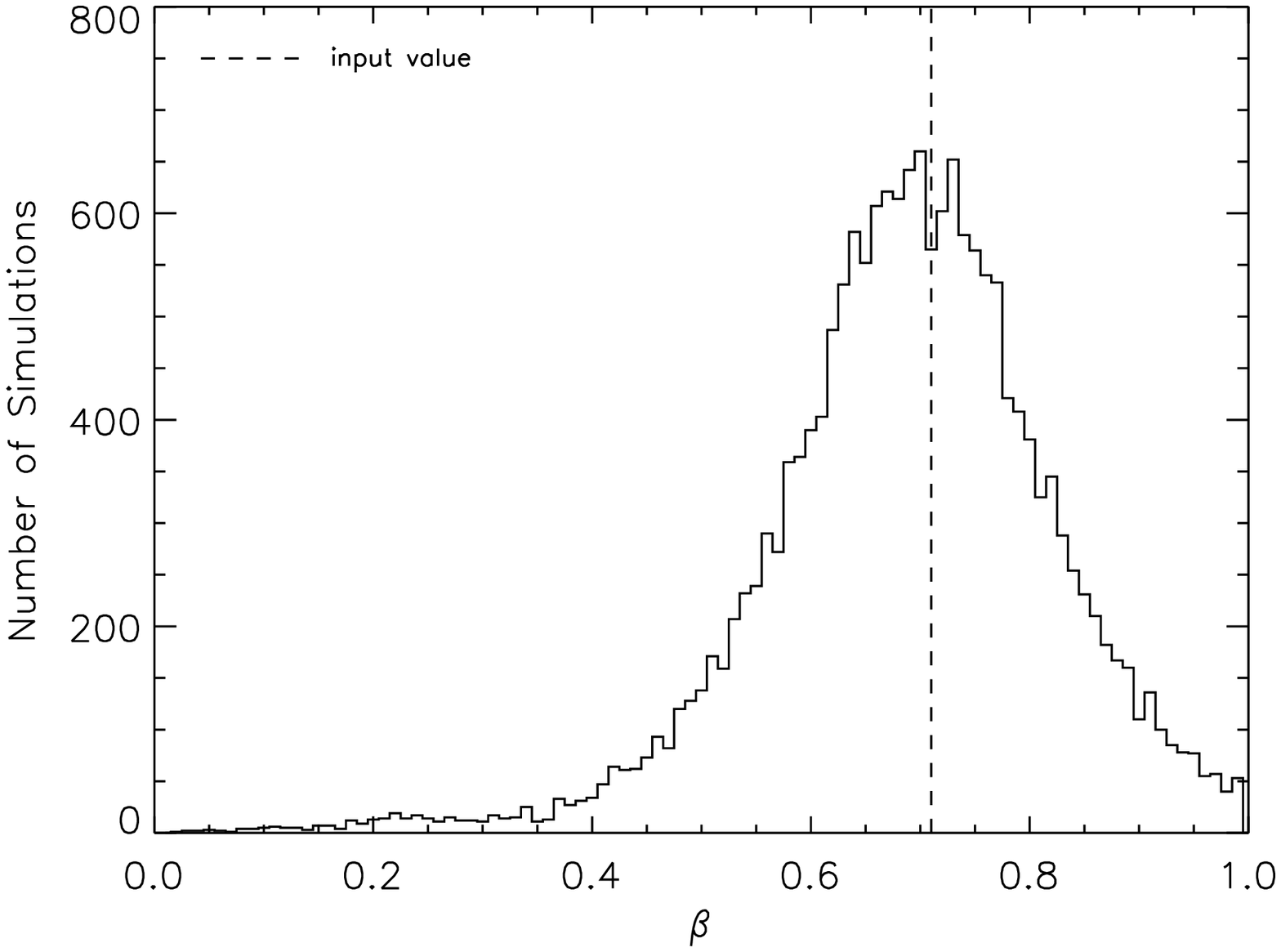}{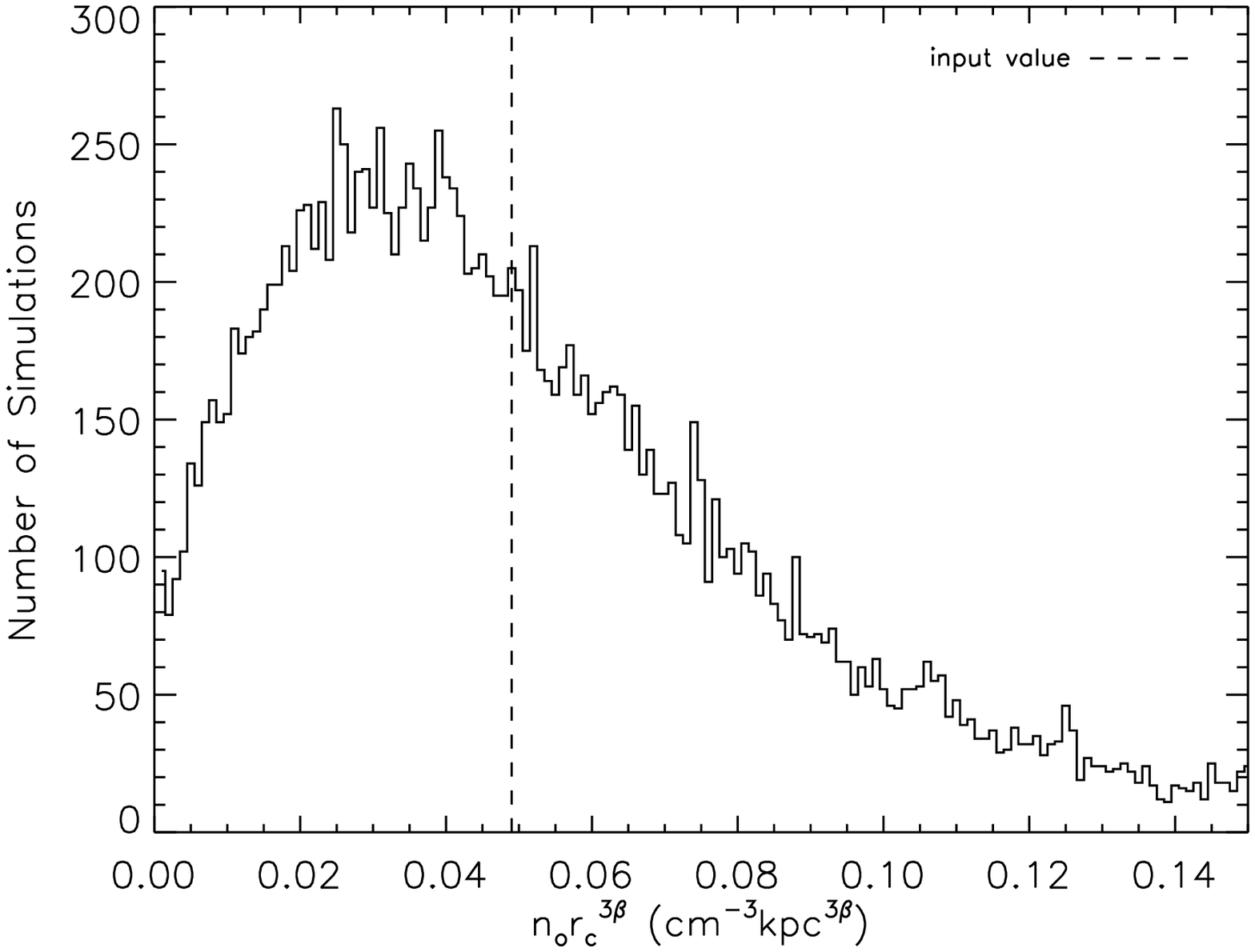}
\includegraphics[height =2.3in, width =5.4in, keepaspectratio=true]{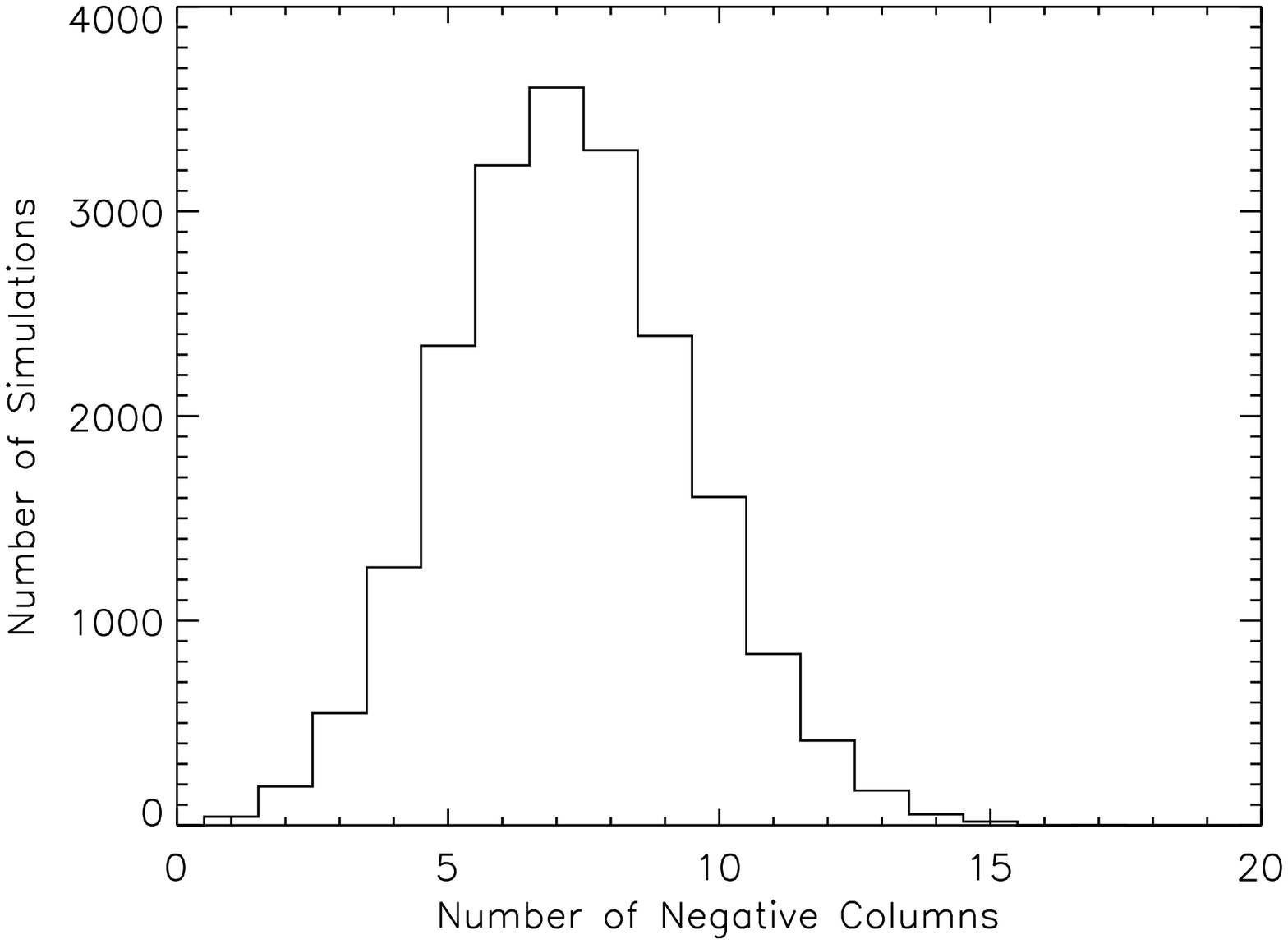}
\caption{Histograms of best-fit parameters $\beta$ and $n_or_c^{3\beta}$ in addition to the number of negative column densities for 20,000 Monte Carlo simulations.  Our input values for each parameter were our best-fit parameters from Table 2.  We found median values of 0.71 for $\beta$ and 0.050 cm$^{-3}$ kpc$^{3\beta}$ for $n_or_c^{3\beta}$, which are consistent with the measured values of 0.71 and 0.048 cm$^{-3}$ kpc$^{3\beta}$ (dashed lines).  The distributions and medians for $\beta$, $n_or_c^{3\beta}$, and other model parameters are consistent with our measured best-fit parameters and their 1$\sigma$ uncertainties, implying we recover our best-fit model with negative column densities in our sample. }

\end{figure}

\begin{figure}

\centering
\includegraphics[height =3.4in, width =8.0in, keepaspectratio=true]{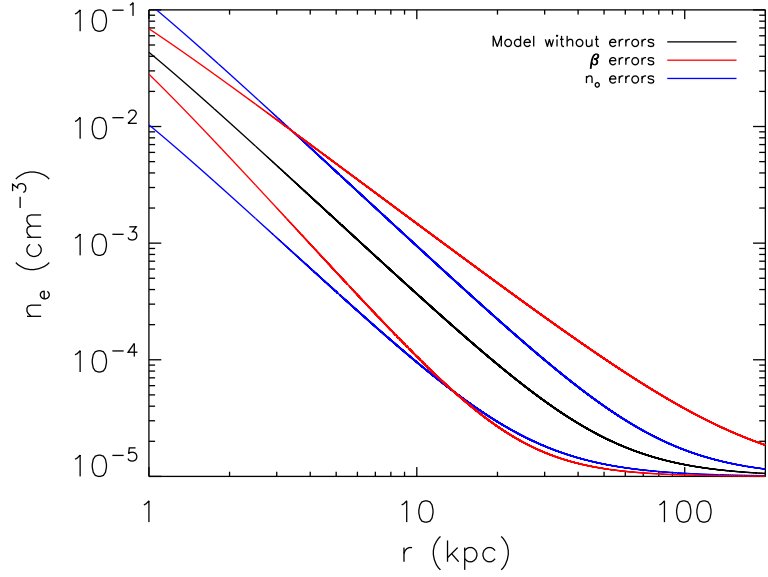}
\caption {Density profile given our best-fit parameters (black line) with 1$\sigma$ errors on $\beta$ (red line) and $n_o$ (blue line).  The profile also includes an additional ambient medium of $n_e$ = 1$\times10^{-5}$ cm$^{-3}$ to account for observed ram-pressure stripping of dwarf spheroidal galaxies.   }

\end{figure}


\begin{figure}

\centering
\includegraphics[height =3.4in, width =8.0in, keepaspectratio=true]{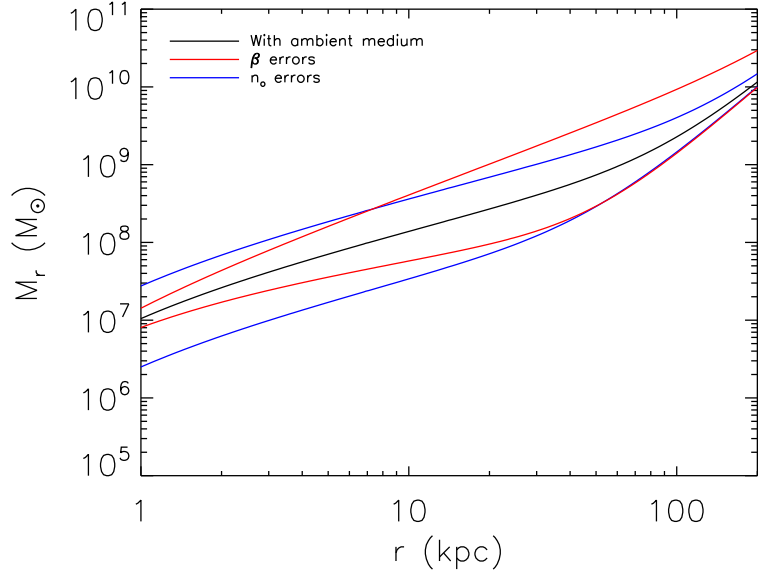}
\caption {Enclosed mass as a function of radius with the same uncertainties as Figure 7.  The enclosed mass of the halo is much smaller than the virial mass of the Milky Way and is only comparable to the stellar + cold gas mass of the Milky Way if the size is comparable to the virial radius of the Milky Way.  The mass profile here is for solar metallicity gas.  }

\end{figure}


\begin{figure}

\centering
\includegraphics[height =3.4in, width =8.0in, keepaspectratio=true]{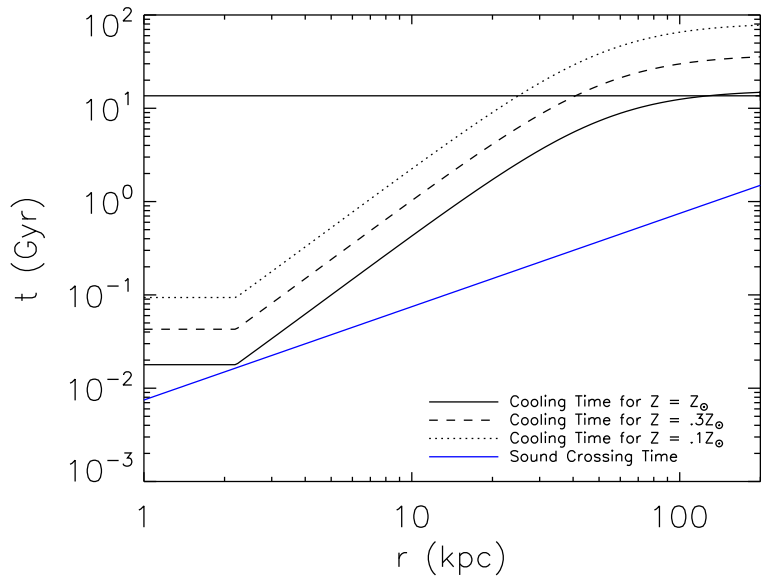}
\caption {Cooling time as a function of radius given our best-fit parameters and assuming the halo is isothermal at log $T$ = 6.1.  The cooling time is sensitive to the assumed metallicity of the halo and is comparable to a Hubble time at the Milky Way's virial radius for solar metallicity gas.  This implies that the halo either has a sub-solar metallicity or is subject to a continuous heating source.  We also plot the sound crossing time as a function of galactocentric radius.  This shows that the cooling time is greater than the sound crossing time at all radii, implying the halo is in hydrostatic equilibrium.}

\end{figure}


\begin{figure}

\centering
\includegraphics[height =3.4in, width =8.0in, keepaspectratio=true]{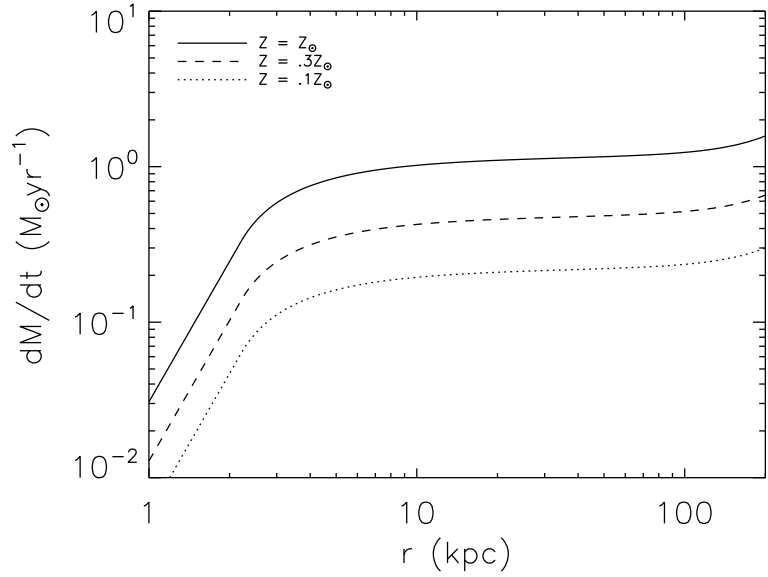}
\caption{Cooling rate (or accretion rate) as a function of radius for our best-fit parameters.  The solar metallicity gas results in an accretion rate that is roughly consistent with the SFR observed in the Milky Way, but sub-solar metallicity gas is more consistent with what has been observed in other spiral galaxies.  }

\end{figure}


\end{document}